\pdfoutput=1
\documentclass[12pt]{iopart}
\usepackage{graphicx}
\usepackage[center]{subfigure}

\newcommand{\deriv}[2]{\ensuremath{\frac{\partial #1}{\partial #2}}}

\begin{document}

\title{Hermes: Global plasma edge fluid turbulence simulations}
\author{B D Dudson and J Leddy}
\address{York Plasma Institute, Department of Physics, University of York, Heslington, York YO10 5DQ, UK}
\ead{benjamin.dudson@york.ac.uk}

\begin{abstract}
The transport of heat and particles in the relatively collisional edge regions of magnetically confined plasmas is a scientifically challenging and technologically important problem. Understanding and predicting this transport requires the self-consistent evolution of plasma fluctuations, global profiles and flows, but the numerical tools capable of doing this in realistic (diverted) geometry are only now being developed.

Here a 5-field reduced 2-fluid plasma model for the study of instabilities and turbulence in magnetised plasmas is presented, built on the BOUT++ framework. This cold ion model allows the evolution of global profiles, electric fields and flows on transport timescales, with flux-driven cross-field transport determined self-consistently by electromagnetic turbulence. Developments in the model formulation and numerical implementation are described, and simulations are performed in poloidally limited and diverted tokamak configurations. 
\end{abstract}

\submitto{\PPCF}

\maketitle

\section{Introduction}
\label{sec:introduction}

The edge of magnetically confined plasmas, such as tokamaks, is where hot confined plasma encounters neutral gas, material surfaces, and the associated impurities. The transport of heat and particles in this region determines the heat loads and erosion rates of plasma facing components (PFCs). Predicting this transport, and exploring means of reducing heat fluxes to PFCs, has played an important role in designing the ITER divertor~\cite{kukushkin2011}, and will be critical to the design of a future DEMO device~\cite{efdaroadmap,wennninger2015}. One of the uncertainties in making these predictions is the transport across the magnetic field, which is not well described by diffusion~\cite{naulin-2007,ghendrih-2012}, and is thought to be turbulent~\cite{dippolito-2011}. Since the fluctuations can be of similar spatial scales and magnitude to average profiles, significant effort has been devoted to developing and testing 3D flux-driven fluid turbulence simulation codes, including GBS~\cite{ricci2012,halpern2016}, TOKAM-3D~\cite{tamain2010} and TOKAM3X~\cite{tamain2016}.

In this paper we present Hermes, a new tool for the simulation of collisional plasmas,
in particular the edge and divertor regions of tokamaks. The ultimate aim of this work is to construct a model capable of
simulating self-consistently the turbulence, plasma profiles and flows in the edge of magnetically confined plasmas over transport timescales.
This is an ambitious undertaking, as it requires the combination of neutral gas and atomic physics, plasma-wall interaction, turbulent transport, and neoclassical effects~\cite{catto-2008}. As a first step towards this, we present in section~\ref{sec:model} a drift-reduced model~\cite{mikhailovskii1984,pfirsch1996,mikhailovskii1997,reiser2012}
which has been constructed based on the derivation of Simakov and Catto~\cite{simakov-2003}, and similar to that derived in~\cite{scott-2003}
but with several modifications to make it more suitable for numerical solution in global geometry. 
This model has good conservation properties, described in section~\ref{sec:conservation},
and has been implemented using the BOUT++ framework~\cite{Dudson2009,dudson2014} using
conservative numerical methods described in section~\ref{sec:numerical}.

Significant advances have been made in improving numerical accuracy and stability. In particular, PETSc~\cite{petsc-web-page,petsc-user-ref,petsc-efficient} has been used to implement a new solver for the axisymmetric potential $\phi$, described in section~\ref{sec:phisolve}. This allows BOUT++ simulations to self-consistently evolve the axisymmetric electric field in X-point geometry for the first time. Results are presented showing that important features of the global equilibrium can be recovered, including the Pfirsch-Schl\"uter current and Geodesic Acoustic Mode (GAM) oscillations~\cite{winsor-1968}. 

As a demonstration of the capabilities of this new model, turbulence simulations are reported in poloidal limiter geometry (the ISTTOK device~\cite{silva-2004}, section~\ref{sec:limiter_turb}), and in X-point geometry (DIII-D, section~\ref{sec:xpt_turb}). Conclusions, limitations of the present model, and future work are discussed in section~\ref{sec:conclusions}.

\section{Hermes model equations}
\label{sec:model}

The plasma model is electromagnetic, and evolves electron density $n_e$, electron pressure $p_e = n_eT_e$ where $T_e$ is the electron temperature, 
parallel ion momentum $n_ev_{||i}$, plasma vorticity $\omega$, and electromagnetic potential $\psi$.
There is no separation between background and fluctuations in this model, since we wish to study regions
such as the tokamak edge where these are of similar magnitude. The model discussed here (Hermes-1) assumes
cold ions; a hot-ion extension of this model (Hermes-2) is currently under development. The evolution equations normalised to the ion gyro-frequency $\Omega_{ci} = eB/m_i$ and sound gyro-radius $\rho_s = \sqrt{eT_e/m_i} / \Omega_{ci}$ are:
\begin{eqnarray}
  \deriv{n_e}{t} &=& -\nabla\cdot\left[n_e\left(\mathbf{V}_{E\times B} + \mathbf{V}_{mag} + \mathbf{b}v_{||e}\right)\right]  \nonumber \\
  && + \nabla\cdot\left(D_\perp\nabla_\perp n_e\right) + S_n
  \label{eq:density}
\end{eqnarray}

\begin{eqnarray}
  \frac{3}{2}\deriv{p_e}{t} &=& -\nabla\cdot\left(\frac{3}{2}p_e\mathbf{V}_{E\times B} + \frac{5}{2}p_e \mathbf{b}v_{||e} + p_e\frac{5}{2}\mathbf{V}_{mag} \right)  \nonumber \\
  && - p_e\nabla\cdot\mathbf{V}_{E\times B} + v_{||e}\partial_{||}p_e  + \nabla_{||}\left(\kappa_{e||}\partial_{||}T_e\right) \nonumber \\
  && + 0.71\nabla_{||}\left(T_e j_{||}\right) - 0.71 j_{||}\partial_{||} T_e + \frac{\nu}{n_e}j_{||}^2  \\
  &&+ \nabla\cdot \left(D_\perp\frac{3}{2} T_e\nabla_\perp n_e\right) + \nabla\cdot\left(\chi_\perp n_e\nabla_\perp T_e\right) + S_p \nonumber
\end{eqnarray}

\begin{eqnarray}
  \deriv{\omega}{t} &=& -\nabla\cdot\left(\omega\mathbf{V}_{E\times B}\right) + \nabla_{||}j_{||} - \nabla\cdot\left(n_e \mathbf{V}_{mag}\right) \nonumber \\
  &&  + \nabla\cdot\left(\mu_\perp\nabla_\perp\omega\right)
  \label{eq:vorticity}
\end{eqnarray}

\begin{eqnarray}
  \frac{\partial}{\partial t}\left(n_ev_{||i}\right) &=& -\nabla\cdot\left[n_ev_{||i}\left(\mathbf{V}_{E\times B} +  \mathbf{b}v_{||i}\right)\right] - \partial_{||}p_e \nonumber\\
  && + \nabla\cdot \left(D_\perp v_{||i}\nabla_\perp n_e\right) - F
  \label{eq:ion_momentum}
\end{eqnarray}

\begin{eqnarray}
  \frac{\partial}{\partial t}\left[\frac{1}{2}\beta_e\psi - \frac{m_e}{m_i}\frac{j_{||}}{n_e}\right] &=& \nu \frac{j_{||}}{n_e} + \partial_{||}\phi - \frac{1}{n_e}\partial_{||} p_e \nonumber \\
 && - 0.71\partial_{||} T_e  \label{eq:ohms_law} \\
&& + \frac{m_e}{m_i}\left(\mathbf{V}_{E\times B} + \mathbf{b}v_{||i}\right)\cdot\nabla\frac{j_{||}}{n_e} \nonumber
\end{eqnarray}
with cross-field E$\times$B and magnetic drifts given by:
\numparts
\begin{eqnarray}
\mathbf{V}_{E\times B} &=& \frac{\mathbf{b}\times\nabla \phi}{B} \\
\mathbf{V}_{mag} &=& -T_e\nabla\times\frac{\mathbf{b}}{B} \label{eq:drifts}
\end{eqnarray}
\endnumparts
Here we use the notation $\nabla_\perp = \nabla - \mathbf{b}\mathbf{b}\cdot\nabla$, $\nabla_\perp^2 = \nabla\cdot\nabla_\perp$, $\partial_{||}f = \mathbf{b}\cdot\nabla f$ and $\nabla_{||}f = \nabla\cdot\left(\mathbf{b} f\right)$. 
The electron beta appearing in equation~\ref{eq:ohms_law} is $\beta_e = \mu_0p_e/B^2$.
The parallel current $j_{||} = \nabla_\perp^2\psi$ is used to calculate the parallel electron velocity using $j_{||} = n_e\left(v_{||i} - v_{||e}\right)$. 
The parallel electron thermal conduction coefficient is the Braginskii value $\kappa_{||e} = 3.2n_ev_{th,e}^2\tau_e$, where $v_{th,e}$ is the electron thermal speed and $\tau_e$ is the electron collision rate, with optional flux limiters as used in SOLPS~\cite{schneider-2006}. 
The resistivity is given by $\nu = \left(1.96\tau_em_i/m_e\right)^{-1}$. 
Anomalous diffusion can be represented in axisymmetric simulations (section~\ref{sec:xpt_fluid}) by particle and thermal diffusivities $D_\perp$ and $\chi_\perp$.

The form of the diamagnetic current, in terms of a magnetic drift $V_{mag}$, is used here because it is more easily
implemented in terms of fluxes through cell faces than the standard form, and hence suitable for the conservative numerical schemes described
in section~\ref{sec:numerical}: The flow speed depends only on the local temperature, rather than on
pressure gradients, and this has been found to improve numerical stability. This approach has also been used in the TOKAMX code~\cite{tamain2016}.

The vorticity is related to the electrostatic potential $\phi$ by
\begin{equation}
\omega = \nabla\cdot\left(\frac{n_0}{B^2}\nabla_\perp \phi\right)
\label{eq:vorticity_definition}
\end{equation}
where the Boussinesq approximation is used, replacing $n_e$ with a constant $n_0$ (with no spatial or temporal variation) in the vorticity equation. The capability to perform simulations without making this approximation has been used in BOUT++ for other models~\cite{angus2014,omotani2016} but is not used here because it cannot yet be used in diverted X-point geometry for the axisymmetric electric field (section~\ref{sec:phisolve}). The calculation of electrostatic potential $\phi$ from the vorticity $\omega$ is a crucial component in all drift-reduced plasma models, including Hermes. Developments to allow equation~\ref{eq:vorticity_definition} to be solved correctly in X-point geometry are described in section~\ref{sec:phisolve}.

\subsection{Conservation properties}
\label{sec:conservation}

The equations~\ref{eq:density}-\ref{eq:ohms_law} are formulated in divergence form, and the operators are cast in terms of fluxes between cells to ensure conservation of the quantity being advected (see section~\ref{sec:numerical}). This is potentially important for advection of plasma density, 
since experience with transport codes has shown that results are sensitive conservation of particle number in high recycling regimes~\cite{kukushkin2011}. 

In order to study conservation of energy, the procedure described in~\cite{scott-2003,scott-2005} is followed: Multiply the vorticity equation~(\ref{eq:vorticity}) by $\phi$, and Ohm's law~(\ref{eq:ohms_law}) by $j_{||}$. Rearranging,
a set of divergence terms and a set of transfer channels are obtained. When integrated over the spatial domain
the divergence terms become fluxes through the boundary, which can be
made to vanish by appropriate choice of boundary conditions. The rate of change of
each form of energy, in the absence of boundary fluxes or sources and neglecting cross-field diffusion and dissipation terms, is given by:
\begin{eqnarray}
  \frac{\partial}{\partial t}\left[\frac{1}{2}\frac{n_0}{B^2}\left|\nabla_\perp\phi\right|^2\right] &=& -\phi\nabla_{||}j_{||} - \phi\nabla\cdot\left(p_e\nabla\times\frac{\mathbf{b}}{B}\right) \label{eq:exb_energy}
\end{eqnarray}
\begin{eqnarray}
  \frac{\partial}{\partial t}\left(\frac{1}{2}nv_{||i}^2\right) &=& -\frac{1}{2}v_{||i}^2\left[\frac{\partial n}{\partial t} + \nabla\cdot\left(n\mathbf{v}_i\right)\right] - v_{||i}\partial_{||}p_e \label{ion_par_energy}
\end{eqnarray}
\begin{eqnarray}
\frac{\partial}{\partial t}\left(\frac{1}{4}\beta_e\left|\nabla_\perp\psi\right|^2 + \frac{m_e}{m_i}\frac{1}{2}\frac{j_{||}^2}{n}\right) &=&  - j_{||}\partial_{||}\phi + v_{||i}\partial_{||}p_e \nonumber \\
 && - \nu \frac{j_{||}^2}{n_e} - v_{||e}\partial_{||}p_e \nonumber \\
 && + 0.71j_{||}\partial_{||}T_e \label{eq:emag_energy}
\end{eqnarray}
\begin{eqnarray}
  \frac{3}{2}\deriv{p_e}{t} &=& -p_e\nabla\times\left(\frac{\mathbf{b}}{B}\right)\cdot\nabla\phi + \nu \frac{j_{||}^2}{n_e} \nonumber \\
  && + v_{||e}\partial_{||}p_e - 0.71 j_{||}\partial_{||} T_e \label{eq:thermal_energy}
\end{eqnarray}
which correspond to the ion $E\times B$ energy (equation~\ref{eq:exb_energy}), ion parallel kinetic energy (equation~\ref{ion_par_energy}), electromagnetic field energy (equation~\ref{eq:emag_energy}), and electron thermal energy (equation~\ref{eq:thermal_energy}).
Each term on the right hand side of equations~\ref{eq:exb_energy}-\ref{eq:thermal_energy} has a corresponding term
in another equation with which it balances, representing a transfer of energy from one form to another.
This model therefore has a conserved energy of the form
\begin{eqnarray}
E = \int dv \Bigg[ &&\frac{m_in_0}{2B^2}\left|\nabla_\perp\phi\right|^2 + \frac{1}{2}m_in_ev_{||i}^2 + \frac{3}{2}p_e + \nonumber \\
  &&\frac{1}{4}\beta_e\left|\nabla_\perp\psi\right|^2 + \frac{m_e}{m_i}\frac{1}{2}\frac{j_{||}^2}{n_e} \Bigg]
\end{eqnarray}
Unfortunately this conservation is not complete in the implementation of this model, because the ion parallel momentum
(equation~\ref{eq:ion_momentum}) does not contain the ion polarisation drift. The first term on the right of
equation~\ref{ion_par_energy} vanishes if the total ion velocity $\mathbf{v}_i$ is used in the advection of ion parallel momentum
equation, but in equation~\ref{eq:ion_momentum} the ion velocity is approximated by the sum of $E\times B$ and parallel flow only
(ion diamagnetic velocity being zero for cold ions). The ion polarisation drift is a higher order than the $E\times B$ and diamagnetic drifts considered here, and including the ion polarisation drift in the ion momentum equation 
would introduce a time derivative into the right hand side, which could not then be solved by the Method of Lines approach adopted by BOUT++. 
This is a challenge of drift-reduced models, though a possible solution is suggested in~\cite{scott-2003} and will be investigated as part of future work. 

\subsection{Boundary conditions}

At the radial boundaries simple Dirichlet or Neumann boundary conditions are used, chosen so that the flux of particles and energy through the boundary goes to zero. A difficulty arises at the core boundary, since the magnetic drift $V_{mag}$ is approximately vertical, and so into the boundary at the top of the device and out of the boundary at the bottom (or vice-versa, depending on the sign of the drift). If this is artificially prevented from convecting particles and thermal energy through the core boundary then a narrow unphysical boundary layer forms. As a result here there can be a net flux of energy through the core boundary. In most cases this source is expected to be small compared to the external input power, but pathological cases could exist. A possible future improvement would be to ensure that the total energy and particle flux integrated over the core boundary vanishes.

In the direction parallel to the magnetic field sheath boundary conditions are needed. The correct boundary conditions to apply at the entrance to the sheath in magnetically confined plasmas is the subject of a long and ongoing debate~\cite{chodura-1982,stangeby-2000,loizu-2012,siddiqui-2015,togo-2016}. Here we adopt relatively simple boundary conditions, leaving the choice of more complex boundary condition to future work. The parallel ion velocity is sonic:
\begin{equation}
v_{||i} \ge c_s 
\end{equation}
where $c_s = \sqrt{eT_e/m_i}$ is the sound speed. In practice the inequality means that if the flow in front of the sheath becomes supersonic
then the boundary condition becomes zero-gradient (Neumann condition). Allowing for this possibility is important, since several mechanisms can produce a supersonic transition,
which have been studied analytically~\cite{ghendrih-2011} and in simulations~\cite{bufferand-2014}.

The parallel current at the sheath is given by:
\begin{equation}
j_{||} = en_e\left[v_{||i} - \frac{c_s}{\sqrt{4\pi}}\exp\left(-\left\{\phi/T_e\right\}\right)\right] \label{eq:sheath_current}
\end{equation}
Note that the ion speed into the sheath is used rather than just $c_s$, to handle the supersonic regime correctly. The electron
flow into the sheath saturates, so in equation~\ref{eq:sheath_current} the ratio $\left\{\phi/T_e\right\}$ is limited to be $\ge 0$. 

The heat flux into the sheath is controlled by equation~\ref{eq:sheath_heatflux}~\cite{stangeby-2000}:
\begin{equation}
q = v_{||i}\left(\frac{1}{2}m_in_ev_{||i}^2 + \frac{5}{2}p_e\right) - \kappa_{||e}\partial_{||}T_e = \gamma_s n_eT_ec_s
\label{eq:sheath_heatflux}
\end{equation}
where here the sheath heat transmission coefficient is taken to be $\gamma_s = 6.5$. Zero gradient (Neumann) boundary condition is used for the electron temperature, so the loss of energy at the sheath is implemented as an additional removal of thermal energy from the final grid cell.

The boundary condition on the electron density at the sheath entrance should be as unrestrictive as possible, to avoid over-constraining the system of equations. Free boundary conditions have been tried, in which the second derivative at the boundary is zero, but zero gradient boundary conditions have been found to be more robust and so are used here.

\subsection{Coordinate system}
\label{sec:coordinates}

A Clebsch coordinate system is used, aligned with the magnetic field such that the equilibrium magnetic field is given by
$\mathbf{B} = \nabla z\times\nabla x$. The $x$ coordinate is a flux coordinate in the radial direction, $z$ is an angular coordinate, and the $y$ coordinate is aligned with the equilibrium magnetic field. In BOUT++ the metric tensor is assumed to be constant in the $z$ direction. 
Two orientations of this coordinate system are used here: For simulations of the large aspect-ratio ISTTOK device in section~\ref{sec:limiter_turb} the $z$ direction is the poloidal angle. For tokamak simulations
in section~\ref{sec:xpt} the standard BOUT/BOUT++ coordinates are used, in which $z$ is the toroidal angle. For long time simulations the large scale magnetic field might be expected to evolve, but this is not currently allowed for in this model: Parallel derivatives are taken along the starting magnetic field, and the deviation due to the electromagnetic potential $\psi$ is assumed to be small. 

\section{Numerical methods}
\label{sec:numerical}

The numerical schemes used to solve the model equations have been chosen for their conservation properties. Some care must be taken over the form of the equations used, since not all analytically equivalent forms have the same numerical properties. Particular care must be taken over the form and choice of numerical method for energy exchange terms, which convert one form of energy to another~\cite{scott-2003}. The shear Alfv\'en wave dynamics along the magnetic field includes exchange of energy between electromagnetic and $E\times B$ energy, so that in the terms
\begin{equation}
\phi\frac{\partial \omega}{\partial t} = \phi\nabla_{||}j_{||}+\ldots \qquad j_{||}\frac{\beta_e}{2}\frac{\partial \psi}{\partial t} = j_{||}\partial_{||}\phi + \ldots
\end{equation}
the operators $\nabla_{||}$ and $\partial_{||}$ should numerically obey the identity
$\phi\nabla_{||}j_{||} = \nabla_{||}\left(\phi j_{||}\right) - j_{||}\partial_{||}\phi$. Fortunately this is quite easy to achieve, and the standard central differencing schemes have this property.
A similar relation exists for the sound wave coupling between parallel ion flow and electron pressure, with the same solution.

The exchange of energy between $E\times B$ energy and electron thermal energy appears as divergence of diamagnetic current in the vorticity equation, and compression of $E\times B$ flow in the pressure equation:
\numparts
\begin{eqnarray}
\phi\frac{\partial \omega}{\partial t} &=& \phi\nabla\cdot\left(p_e\nabla\times\frac{\mathbf{b}}{B}\right) + \ldots \label{eq:diamag_current} \\
\frac{3}{2}\frac{\partial p_e}{\partial t} &=& p_e\nabla\cdot\left(\frac{\mathbf{b}\times\nabla\phi}{B}\right) + \ldots \\
  &=& p_e\left(\nabla\times\frac{\mathbf{b}}{B}\right)\cdot\nabla\phi + \ldots \label{eq:exbcompress}
\end{eqnarray}
\endnumparts
Here the second form of the $E\times B$ compression term is used (equation~\ref{eq:exbcompress}) and both terms discretised with central differences since this then ensures that these terms combine into a divergence without relying on the properties of $\nabla\times\frac{\mathbf{b}}{B}$, for example its divergence in the numerical scheme.

\begin{figure}[h]
\centering
\includegraphics[width=0.5\columnwidth]{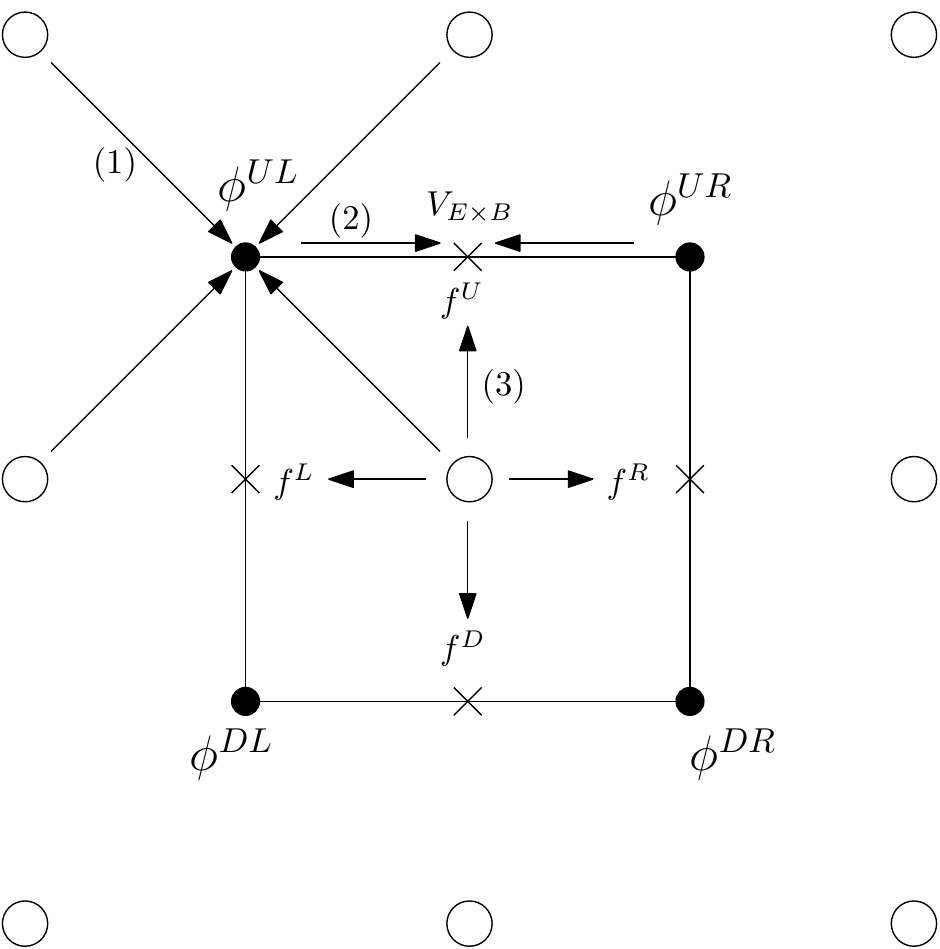}
\caption{Advection of $f$ by potential $\phi$, both defined at cell centres (open circles): (1) $\phi$ is interpolated onto cell corners (solid circles)
(2) Derivatives are taken along cell boundaries to calculate velocity (crosses)
(3) The values of $f$ at both sides of the boundary is constructed. The flux is calculated using $f$ from the upwind side
of the boundary.}
\label{fig:advect_fv}
\end{figure}

The $E\times B$ advection terms are discretised using the scheme
illustrated in figure~\ref{fig:advect_fv}. The potential is first interpolated onto cell corners,
in order to preserve a divergence-free flow in the absence of magnetic curvature. The velocity on each cell boundary is then calculated by taking the derivative of the potential
along the boundary. Note that this (along with the interpolation) makes the method at best 2$^{nd}$-order accurate.
The value of the field $f$ at either side of the boundary is determined by the choice of numerical scheme. This is done
in each dimension independently, so each scheme must construct the values of $f$ at the left ($f^L$) and right ($f^R$) boundaries.
The finite volume schemes implemented to calculate these values include 2$^{nd}$-order Fromm and 4$^{th}$-order XPPM methods~\cite{peterson2013}. It has been found that use of the XPPM method results in a slow convergence of the implicit time integration
scheme usually used (CVODE from the SUNDIALS suite~\cite{hindmarsh2005}), and so here the Fromm method is used for advection by $E\times B$ and magnetic drifts.

\subsection{Poloidal flows}
\label{sec:poloidal_flows}

Poloidal flows due to $E\times B$ drift are important for the Geodesic Acoustic Mode (GAM) oscillation~\cite{winsor-1968,zhou-2015}, and can play a role in the experimentally observed edge asymmetries and flow patterns~\cite{rognlien-1999,chankin-2000,chankin-2015}.
In the tokamak coordinate system used here, the poloidal coordinate is transformed into the coordinate parallel to the magnetic field. The usual way to represent poloidal flows is to split the $E\times B$ flow into two pieces:
\begin{equation}
\nabla\cdot\left(n_e\frac{\mathbf{b}\times\nabla\phi}{B}\right) = \frac{\mathbf{b}\times\nabla\phi}{B}\cdot \nabla n_e + n_e \left[\nabla\times\left(\frac{\mathbf{b}}{B}\right)\right] \cdot\nabla\phi
\label{eq:exb_split}
\end{equation}
The first term is then represented as a Poisson bracket, whilst the
compressional terms leading to GAM oscillations are contained in the second term. The advantages of this approach are that each of these terms can be discretised using the Arakawa method~\cite{arakawa-1966} which has minimal dissipation and respects the symmetries of the underlying equations. 
Unfortunately in general geometry it is extremely difficult to ensure that the resulting numerical method conserves particles since these terms must combine so as to obey the analytic integral relation. 
Here we adopt the approach also used in~\cite{tamain2016}, and write the advection in flux-conservative form:
\numparts
\begin{eqnarray}
\nabla\cdot\left(n_e\frac{\mathbf{b}\times\nabla\phi}{B}\right) &=& \frac{1}{J}\frac{\partial}{\partial\psi}\left(Jn_e\frac{\partial\phi}{\partial z} \right) - \frac{1}{J}\frac{\partial}{\partial z}\left(Jn_e\frac{\partial\phi}{\partial\psi}\right)  \\
                                                              &+& \frac{1}{J}\frac{\partial}{\partial\psi}\left(Jn_e\frac{g^{\psi\psi}g^{yz}}{B^2}\frac{\partial\phi}{\partial y}\right) \\
&-& \frac{1}{J}\frac{\partial}{\partial y}\left(Jn_e\frac{g^{\psi\psi}g^{yz}}{B^2}\frac{\partial\phi}{\partial\psi}\right)
\end{eqnarray}
\endnumparts
The first two terms describe the drift-plane motion, which is calculated
using the scheme illustrated in figure~\ref{fig:advect_fv}. The third term
describes radial flow due to poloidal electric fields, whilst the fourth (last) term describes poloidal flow due to radial electric fields. These poloidal flows are more difficult to implement than the drift-plane motion, particularly in the vicinity of an X-point where the cell corner is shared between eight cells rather than the usual four. Here we adopt a simpler scheme for the poloidal flows, in which 
the gradients of the potential $\phi$ are first calculated at cell centers, then interpolated to the cell boundaries to calculate the flux.

\subsection{Numerical dissipation}

The numerical methods employed here use collocated central differencing, and so some form of numerical dissipation is necessary to control zig-zag/chequerboard modes on the grid scale. This
numerical dissipation must be carefully chosen so as not to introduce unphysical modes or instabilities. Perpendicular to the magnetic field we use classical and anomalous perpendicular diffusion of density, pressure, and vorticity ($D_\perp$, $\chi_\perp$ and $\mu_\perp$ in equations~\ref{eq:density}-\ref{eq:ohms_law}). In the direction parallel to the magnetic field we use a combination of 4$^{th}$-order dissipation operators, corresponding to hyperviscosity terms in the flow variables ($v_e$, $v_i$), and Added Dissipation~\cite{murthy-2002} in the scalar variables ($n$, $p_e$, $\omega$). Added Dissipation is implemented as a flow between cells which is driven by the third derivative of the plasma pressure at the cell boundary. This conserves the flux of the advected quantity (density, thermal energy, momentum) whilst suppressing grid-scale oscillations due to the collocated numerical scheme.

\section{Poloidal limited tokamak geometry}
\label{sec:limiter_turb}

\begin{figure}
\centering
\includegraphics[width=0.5\columnwidth]{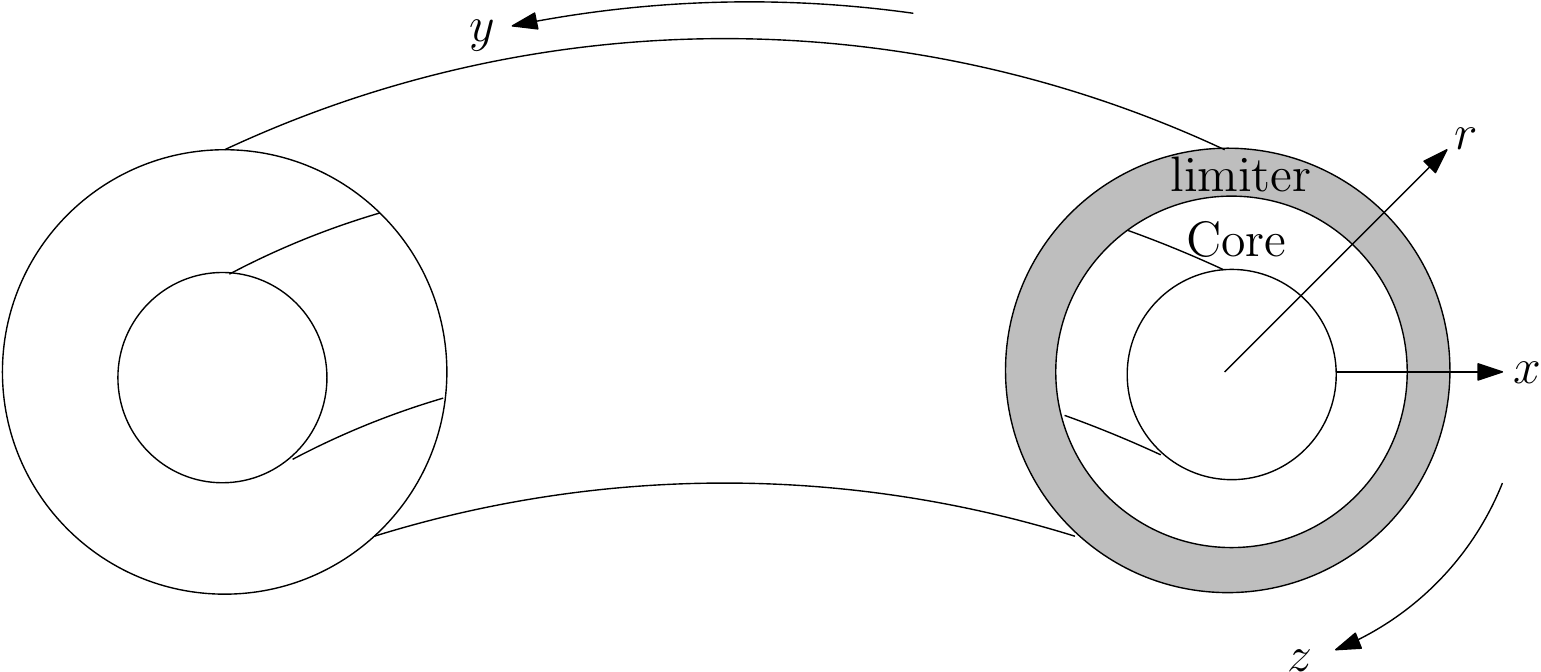}
\caption{Coordinate system used in the ISTTOK simulations. A poloidal limiter is located at one toroidal location, and
the radial domain consists of a region of closed magnetic flux surfaces surrounded by a SOL in which field-lines intersect the limiter.}
\label{fig:isttok}
\end{figure}

We first simulate turbulence in the ISTTOK device~\cite{silva-2004}, a large aspect-ratio, poloidally limited tokamak. Major radius is
$R\simeq 46$cm, minor radius $r\simeq 8.5$cm, toroidal magnetic field $B_\zeta\simeq 0.5$T, and safety factor $q\sim 5-7$. 
In order to accomodate the poloidal limiter, we use a coordinate system in which the $z$ direction is aligned with the poloidal direction $\theta$, shifted so that the toroidal angle $\zeta$ becomes identified with the direction along the magnetic field, $y$:
\begin{equation}
x = \psi \qquad y = \zeta \qquad z = \theta - \zeta/q
\end{equation}
This coordinate system is illustrated in figure~\ref{fig:isttok}. In the core region the magnetic flux-surfaces are closed, so where the $y$ domain connects onto itself a twist-shift boundary using FFTs in the $z$ direction is used to map one end of the flux-tube to the other. This method, and the neglect of metric tensor variation in the poloidal ($z$) direction are possible only due to the large aspect ratio, so this coordinate system cannot yet be used in realistic X-point geometry (section~\ref{sec:xpt_turb}).

In order to maintain a quasi-steady state profile, sources of heat and particles are needed close to the inner boundary. To achieve a specified core density and temperature, we use a Proportional-Integral (PI) feedback controller on the heating and particle sources in the core region. These sources are poloidally uniform, and are limited so that they can only be positive, preventing unphysical removal of heat or particles. 

A simulation has been run with 68 radial points, 16 toroidal, and 256 poloidal points. This corresponds to a resolution of $0.3$mm in the radial direction, and $2$mm in the poloidal direction. The typical turbulence length scale is a multiple of $\rho_s\simeq 0.6$mm at $T_e = 5eV$, so these simulations are probably not fully resolved. Higher resolution simulations will be required in order to carry out a quantitative validation exercise.

\begin{figure}
\centering
\subfigure[Electron pressure $p_e$, ballooning on the outboard (bad curvature) side.]{
  \label{fig:isttok-pxz}
  \includegraphics[width=0.45\columnwidth]{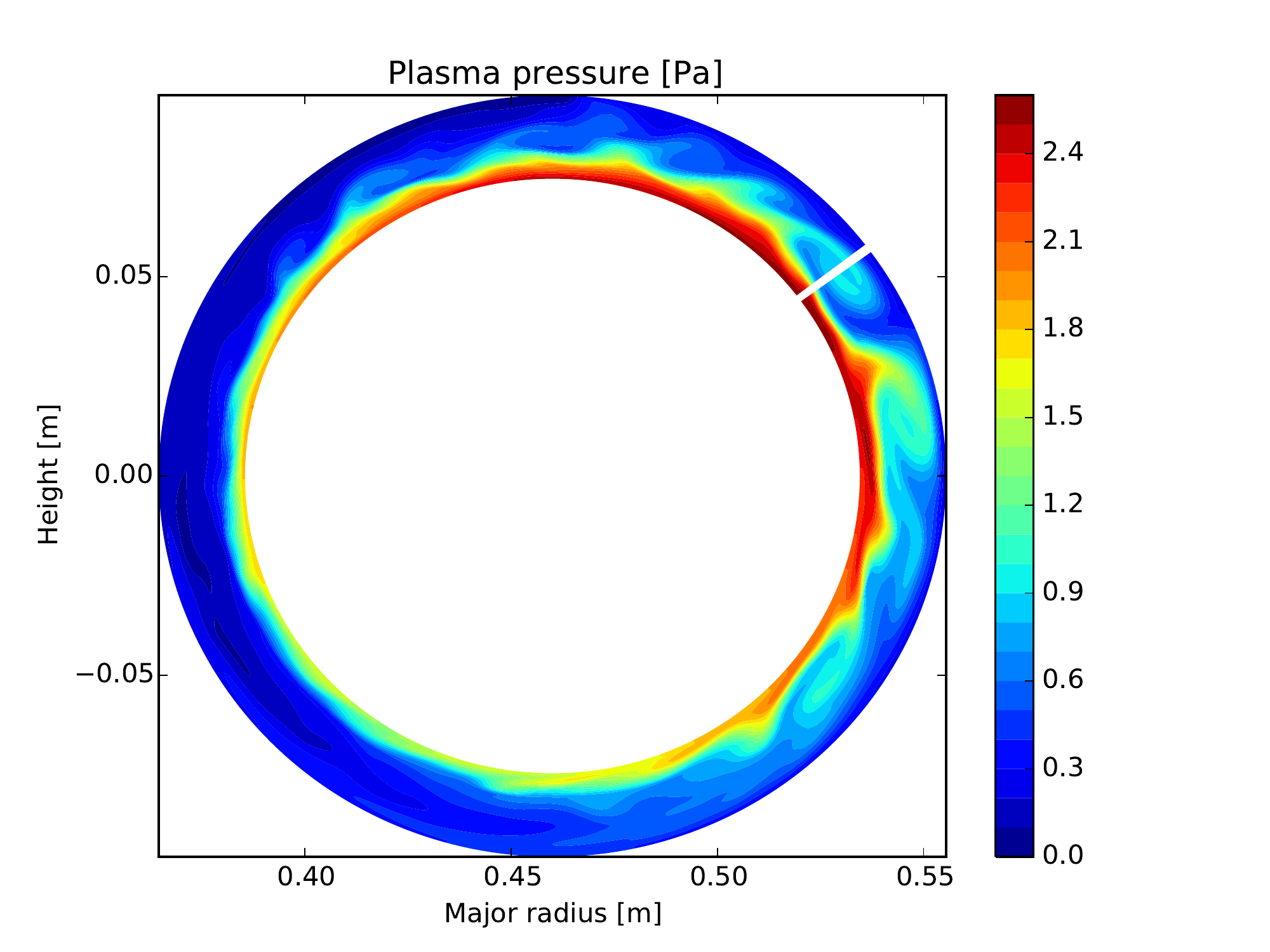}
}
\subfigure[Current density. The equilibrium Pfirsch-Schl\"uter current can be seen in the closed field-line region]{
  \label{fig:isttok-jpar}
  \includegraphics[width=0.45\columnwidth]{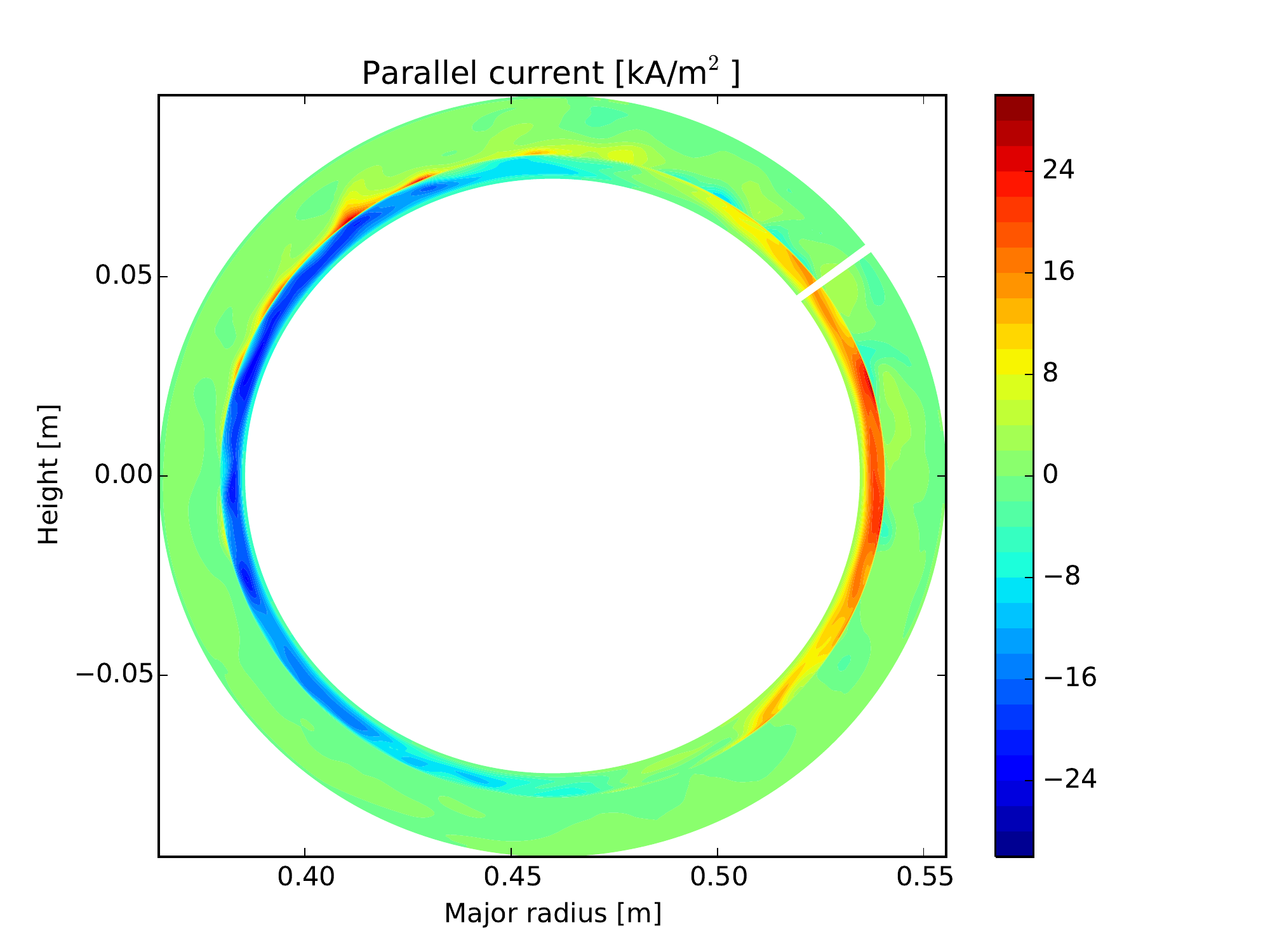}
}
\caption{Pressure and current density at a fixed time and fixed toroidal angle, showing
poloidal asymmetry in the pressure, and large fluctuations in all quantities}
\end{figure}

\begin{figure}
\centering
\subfigure[Inside LCFS]{
  \label{fig:isttok-core}
  \includegraphics[width=0.45\columnwidth]{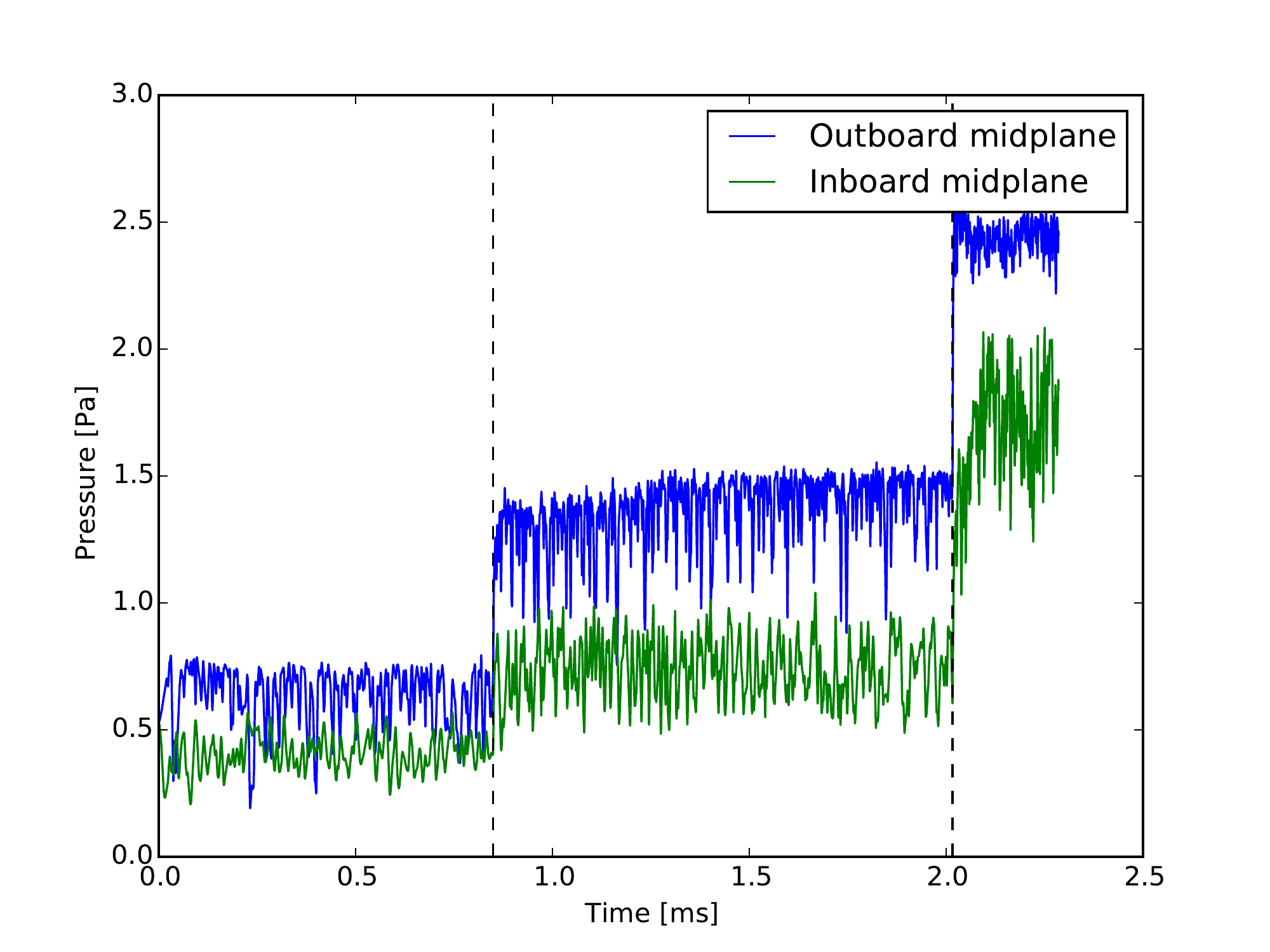}
}
\subfigure[Outside LCFS]{
  \label{fig:isttok-sol}
  \includegraphics[width=0.45\columnwidth]{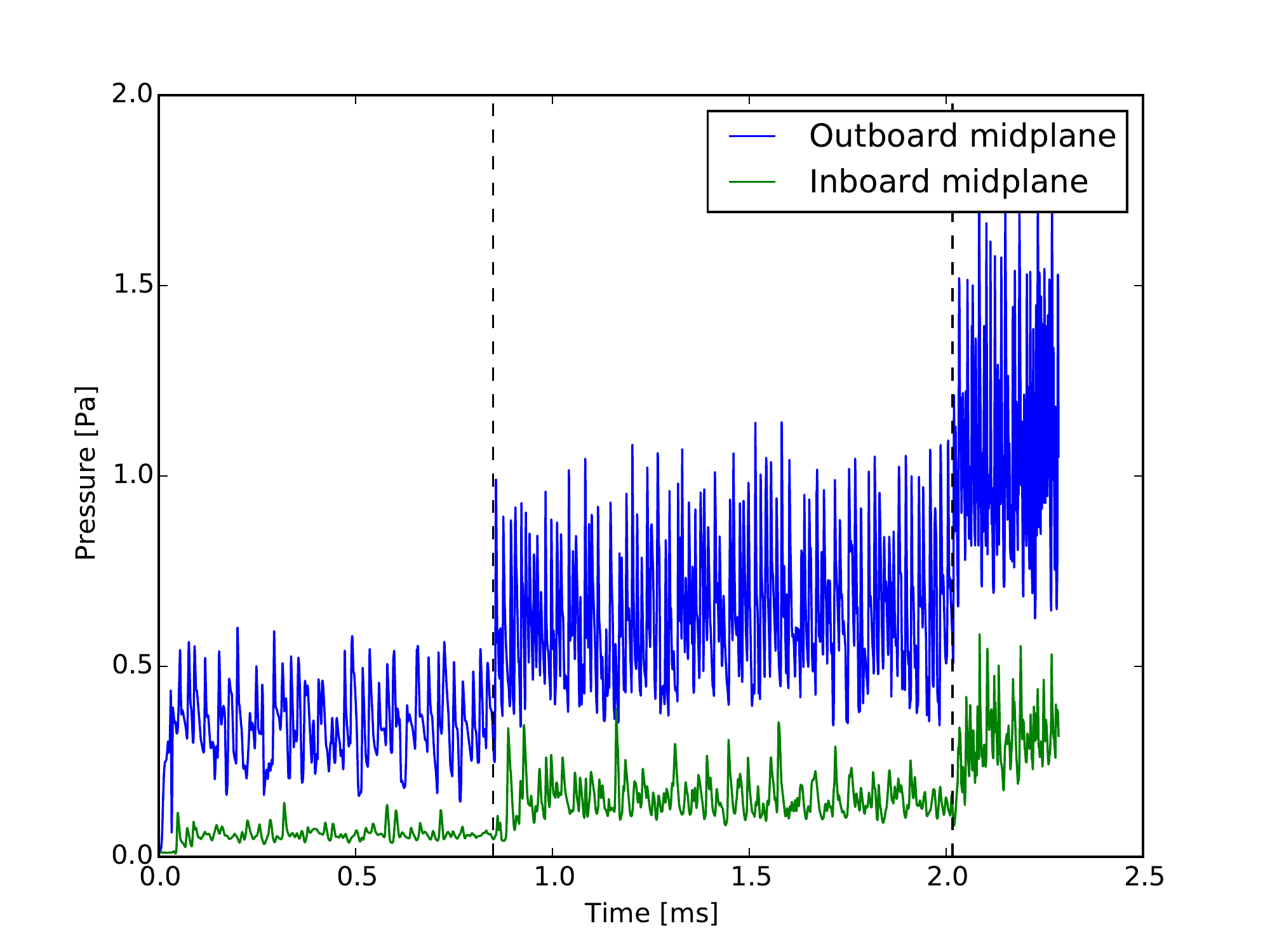}
}
\caption{Fluctuations in electron pressure as a function of time. Vertical dashed lines mark changes in
the edge temperature, from 5eV to 10eV and then to 20eV.}
\label{fig:isttok-time}
\end{figure}

The plasma profiles develop along with the turbulence, and are not prescribed. A
snapshot of the electron pressure is shown in figure~\ref{fig:isttok-pxz}, showing
a poloidal asymmetry. The pressure
gradient gives rise to a Pfirsch-Schl\"uter current,
which can be seen in figure~\ref{fig:isttok-jpar}. Both pressure and parallel current
profiles contain large fluctuations, of the same order as the time averaged value,
underlining the importance of treating the background and fluctuations on the same footing
in the plasma edge.
Fluctuations in plasma pressure as a function of time are shown in figure~\ref{fig:isttok-time}, in which the input power was increased at around $t=0.8$ms and $t=2.0$ms.
Differences between the fluctuations in the closed field-line region inside the Last
Closed Flux Surface (LCFS), and those outside the LCFS can be seen: Inside the LCFS (figure~\ref{fig:isttok-core}) fluctuations are of similar amplitude on the inboard and outboard side,
whilst outside the LCFS (figure~\ref{fig:isttok-sol}) fluctuations are larger on the outboard side, and a clear difference can be seen in the average pressure between the inboard and outboard side. This difference in average pressure between the inboard and outboard sides can be
maintained in ISTTOK partly because these regions are not connected by parallel transport outside the LCFS, since field-lines intersect the poloidal limiter after travelling $1/q \sim 0.2$ of a poloidal circuit.
More detailed study with a higher resolution grid, and comparison against experiment, will be the subject of future work.

\section{Tokamak X-point geometry}
\label{sec:xpt}

Hermes allows simulation of both axisymmetric transport (section~\ref{sec:xpt_fluid}) and 3D turbulence (section~\ref{sec:xpt_turb}) in tokamak X-point geometry. Here we describe the simulation procedure, and some features of the results. As with the ISTTOK asimulations, more detailed analysis is left to a more specific future publication. 

The resolution of these simulations is ($48\times 128 \times 128$) points in the (radial, parallel, toroidal) directions. Due to the field-aligned coordinate system, the effective poloidal resolution is much higher than this, being determined by the pitch of the magnetic field lines~\cite{Dudson2009}.

\subsection{Coordinates}

For X-point tokamak simulations the standard BOUT/BOUT++ coordinates are used. 
In terms of orthogonal toroidal coordinates $\left(\psi, \theta, \zeta\right)$ these are~\cite{xu-2008,Dudson2009}
\begin{equation}
x = \psi  \qquad y = \theta \qquad z = \zeta - \int_{\theta_0}^\theta\frac{B_\phi h_\theta}{B_\theta R}d\theta
\label{eq:coordinates}
\end{equation}
where $h_\theta$ is the poloidal arc length per radian (minor radius for circular cross-section), $R$ the major radius, $B_\phi$ the toroidal magnetic field and $B_\theta$ the poloidal magnetic field. As described in \cite{Dudson2009}, the shifted metric method~\cite{dimits-1993,scott01} is used to reduce cell deformation due to magnetic shear.

\subsection{Fluid transport}
\label{sec:xpt_fluid}

As discussed in section~\ref{sec:introduction}, 3D plasma turbulence on transport timescales is challenging. This can be made more difficult by the simulation starting conditions, which may be far from an equilibrium solution to the evolving equations~\ref{eq:density}-\ref{eq:ohms_law}, resulting in transient axisymmetric oscillations which are time consuming to evolve through. In order to reach a quasi-steady state more quickly, we first evolve only the 2D axisymmetric transport equations, without plasma currents or electric fields, using imposed cross-field anomalous transport coefficients in the spirit of 2D transport codes such as SOLPS~\cite{schneider-2006}, UEDGE~\cite{rognlien-1999,rognlien-2002} or EDGE2D~\cite{chankin-2000}. 

\begin{figure}[htbp!]
\centering
\includegraphics[width=0.7\columnwidth]{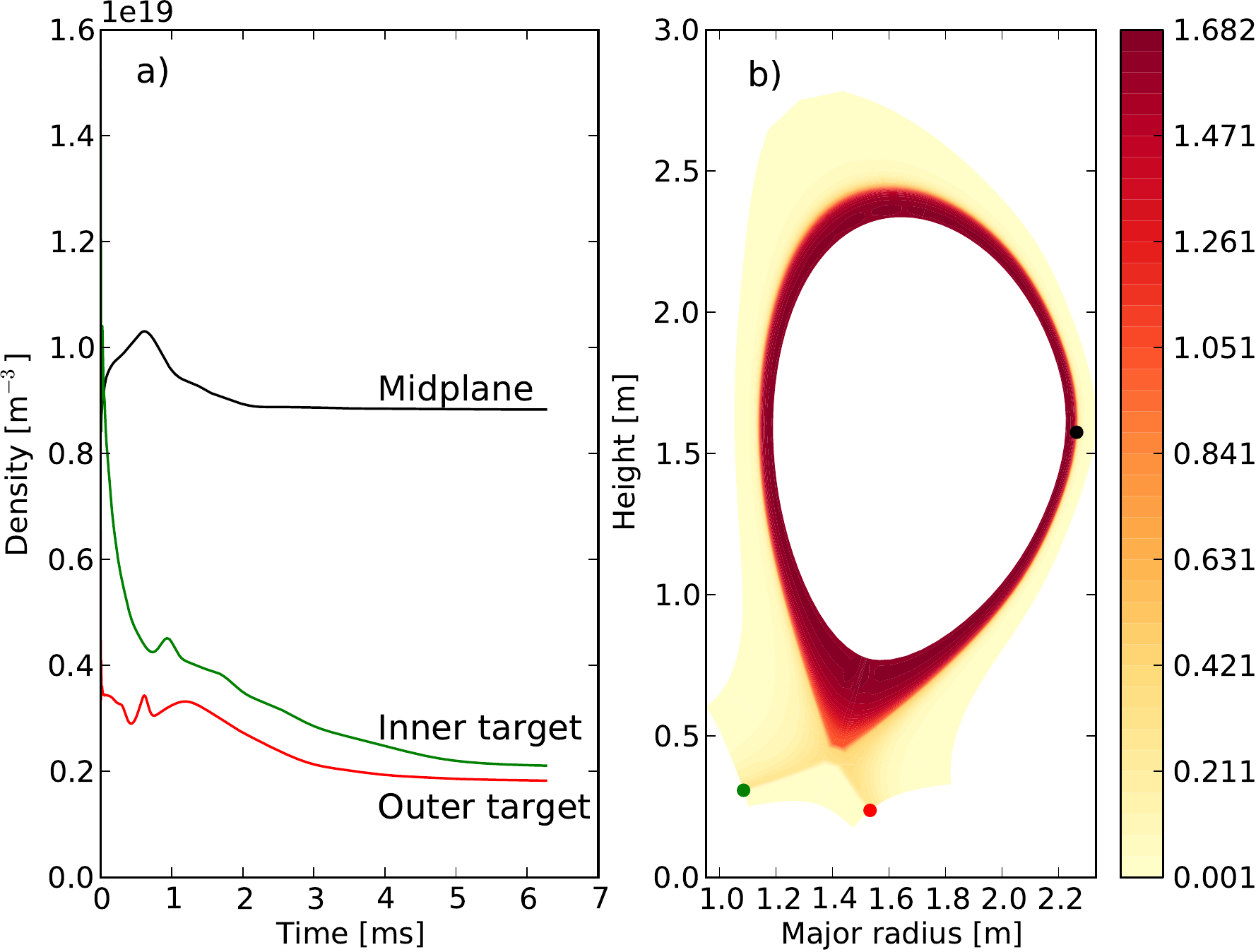}
\caption{Evolution of electron density in a fluid transport simulation as a function of time (a) and the final state (b).}
\label{fig:d3d_fluid_ne}
\end{figure}

Using spatially uniform anomalous diffusion coefficients $D=0.1$m$^2$/s and $\chi=0.2$m$^2$/s the result is shown in figure~\ref{fig:d3d_fluid_ne}.
Since there is no recycling of plasma at the target plates, the density falls from midplane to divertor, and is lower in the divertor than would be typical experimentally. The inclusion of neutral gas and plasma recycling is beyond the scope of this paper, but will be reported elsewhere. As with the ISTTOK simulations, a PI feedback controller is used to control particle and power source, so as to achieve the required core density and temperature.

\subsection{Solution to $\phi$ in X-point geometry}
\label{sec:phisolve}

Once the fluid transport simulation has reached quasi-steady state, we then allow evolution of the vorticity equation and parallel Ohm's law, still in 2D (axisymmetric) mode.

The calculation of electric potential $\phi$ in this model 
requires inverting the elliptic equation~\ref{eq:vorticity_definition},
which is a 2D problem on a curved surface embedded in the 3D domain.
The operator is written in divergence form as
\begin{equation}
\nabla\cdot\left(\frac{n_0}{B^2}\nabla_\perp \phi\right) = \frac{1}{J}\frac{\partial}{\partial u^i}\left(J\frac{n_0}{B^2} g^{ij}\left(\nabla_\perp\phi\right)_j\right)
\label{eq:vort_div_form}
\end{equation}
The metric components $g^{ij}$ which couple the toroidal ($z$) and parallel projection of the poloidal component ($y$) are non-zero, so this is a 3D operator in the field-aligned coordinates used here (equation~\ref{eq:coordinates}).

The technique used in most previous BOUT~\cite{xu-2008} and BOUT++~\cite{Dudson2009} simulations was to neglect derivatives along the magnetic field, being small relative to the $x$ and $z$ derivatives in the ordering used ($k_{||}\ll k_\perp$),
and Fourier transform in the toroidal direction, this being possible when the Boussinesq approximation
is used. This reduces the problem to a set of 1D equations in radial coordinate $x$, which are then solved
using the direct Thomas algorithm for tridiagonal systems. This method is computationally efficient, but
for the $n=0$ mode the magnitude of the diagonal elements in this tridiagonal matrix becomes equal
to the sum of the magnitudes of the off-diagonal elements, and the matrix is not strictly diagonally dominant. 
By comparing this solver with an iterative method, it was found that despite the relative error in the $n=0$ component
on a single solve being of the order of $10^{-6}$, the iterative solver did not result in the growth of a numerical instability
in cases where the direct solver did. By varying the tolerances in the iterative solver, it has been verified that this is not due
to an effective smoothing error in the iterative solver. We therefore conclude that the direct solver
is responsible, and should not be used for the $n=0$ component.

\begin{figure}[htbp!]
\centering
  \includegraphics[width=0.7\columnwidth]{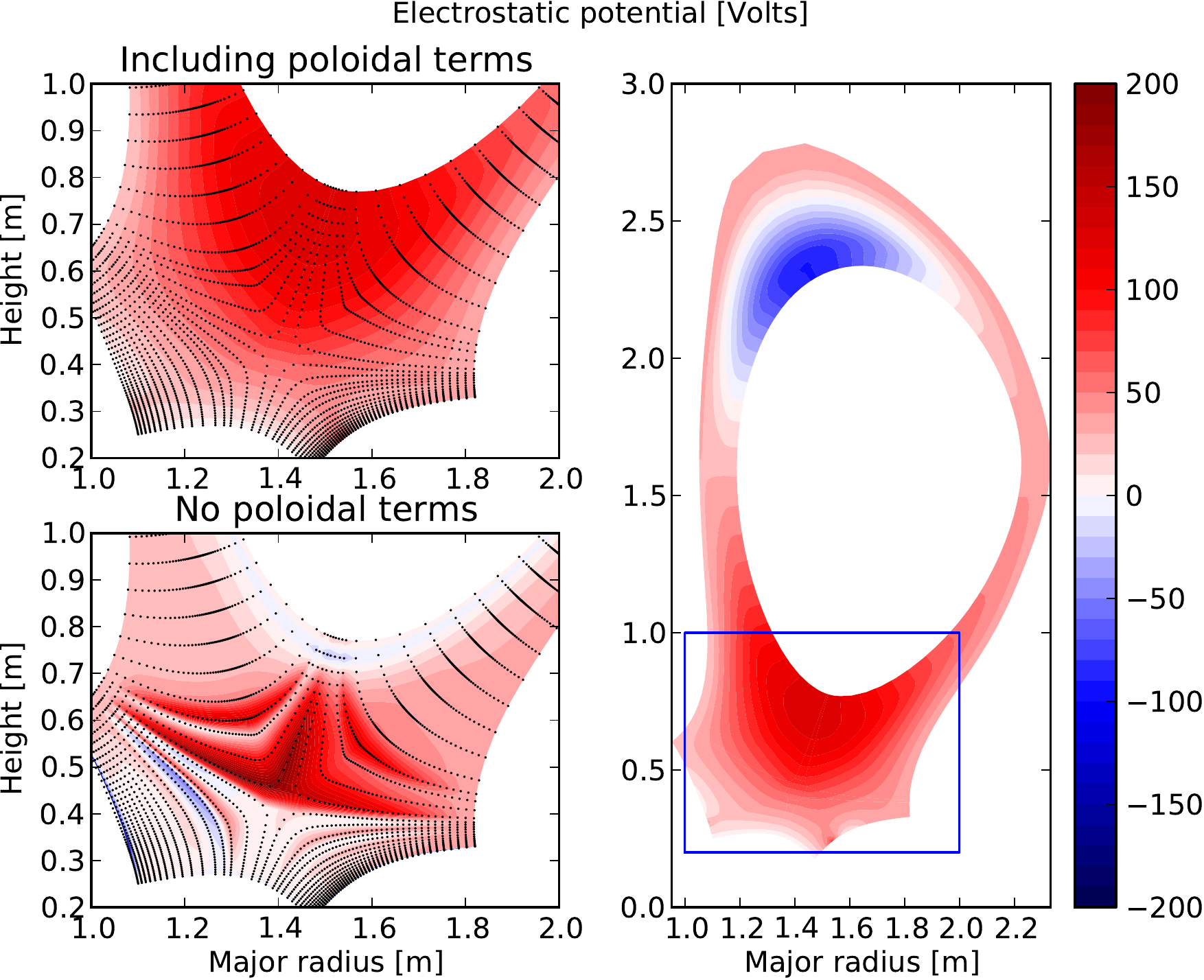}
\caption{Axisymmetric electrostatic potential in X-point geometry, calculated with or without retaining poloidal ($y$) derivatives. 
The vorticity $\omega$ is the same in both cases.}
  \label{fig:d3d_phi}
\end{figure}

In addition to the issues discussed above, a further correction has been identified for simulations
in X-point geometry: The flute ordering ($k_{||}\ll k_\perp$) used to justify dropping parallel
derivatives is not sufficient close to the X-point, particularly for low toroidal mode numbers.
Neglecting parallel (poloidal) derivatives does not on its own cause numerical instability,
since it does not result in a spurious source of energy,
but does significantly alter the solution, and introduces sharp gradients which can in turn lead
to numerical problems in other terms. 
This is illustrated in figure~\ref{fig:d3d_phi}, which shows the potential at a single time
early in the development of the axisymmetric electrostatic potential in X-point geometry. Neglecting parallel derivatives
in this coordinate system decouples grid points in the poloidal direction, in particular the
coordinate lines passing either side of the X-point. Close to the X-point (marked with a box),
this produces a discontinuity in the poloidal direction. This discontinuity is unphysical,
and is resolved by retaining the $y$ derivatives in equation~\ref{eq:vort_div_form}. 
 This new solver does not yet handle finite toroidal mode numbers, so $n>0$ modes are inverted using the same solver as previously used in BOUT++, making the flute approximation to neglect parallel (poloidal) derivatives. Since this simplification is unlikely to be accurate for low $n$ modes in X-point geometry, here we retain only $n=0$ and $n \ge 4$ modes, removing $n=1,2,3$ by simulating only $1/4$ of the torus. 
Note that this decomposition in toroidal modes is only possible
here because we have made the Boussinesq approximation in equation~\ref{eq:vorticity_definition}. If this approximation is not used then a 3D solution of $\phi$ becomes necessary in this coordinate system. An efficient means to do this is currently under development, and will be reported elsewhere.

\begin{figure}[htbp!]
\centering
\includegraphics[width=0.6\columnwidth]{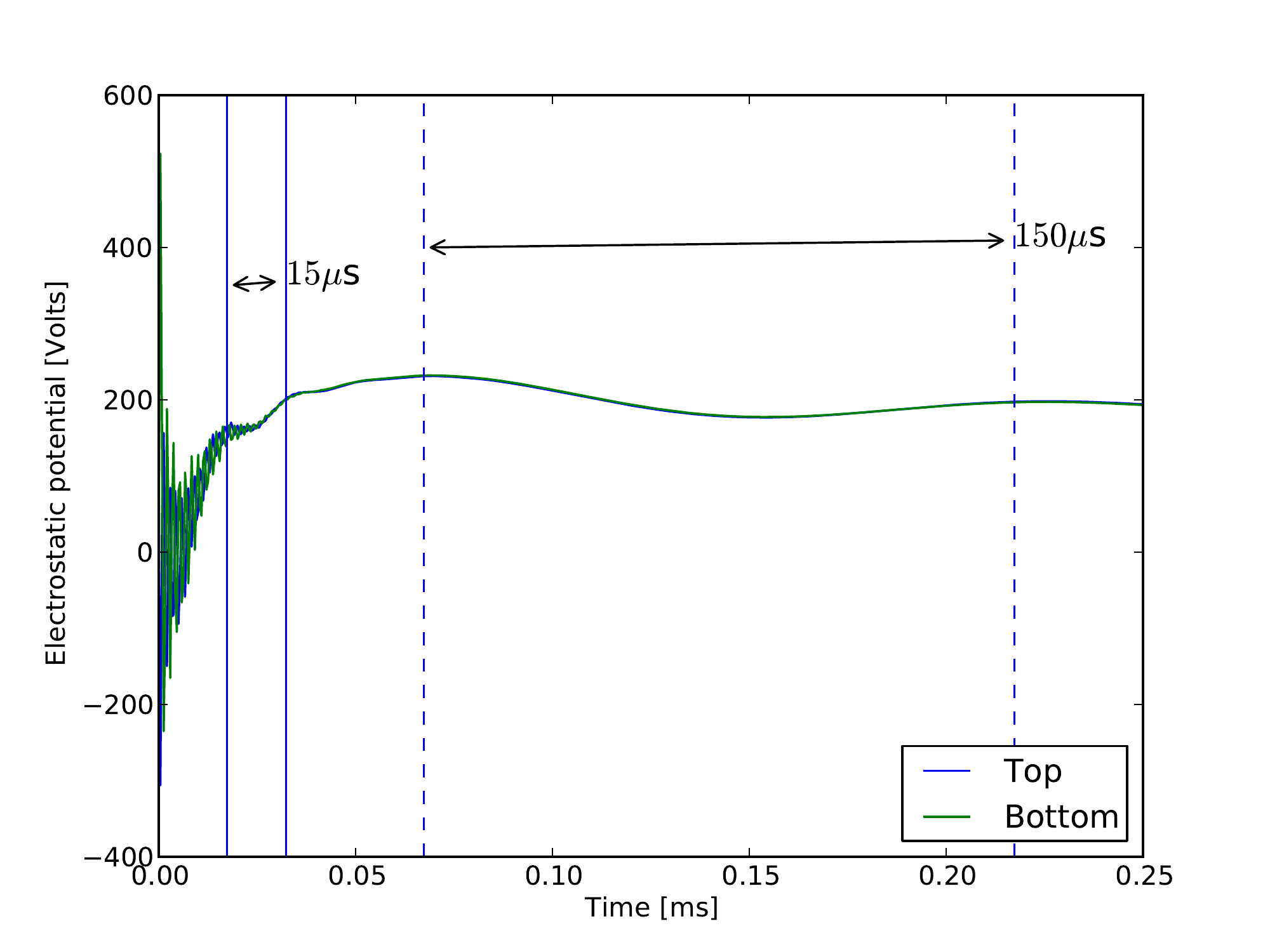}
\caption{Evolution of the electrostatic potential $\phi$ on the same flux surface at the top and bottom (near X-point) of the plasma. Starts from an axisymmetric fluid solution (fig~\ref{fig:d3d_fluid_ne}) with zero electric field and zero parallel current.}
\label{fig:d3d_axisym_phi}
\end{figure}

Using this new axisymmetric field solver, the resulting
evolution of the electrostatic potential $\phi$ in the core region
is shown for two poloidal locations on the same flux surface in figure~\ref{fig:d3d_axisym_phi}.
Three oscillation frequencies are apparent: (1) An oscillation during the first $20\mu$s with a frequency of approximately $500$kHz, in which the potentials at the top and bottom of the plasma are out of phase; (2) A strongly damped oscillation with frequency around $67$kHz during the first $\sim 50\mu$s, in which the potentials on the same flux surface are approximately in phase; and (3) a much slower oscillation with frequency around $6.7$kHz, with potentials in phase. The simple analytical large aspect-ratio estimates of local wave frequencies are: shear Alfv\'en wave $f_A = v_A / \left(2\pi R q\right) \simeq 550-1100$kHz; GAM frequency $f_{GAM} = \frac{c_s}{2\pi R}\sqrt{2 + 1/q^2} \simeq 3-11$kHz; and the parallel sound wave $f_s = c_s/\left(2\pi R q\right) \simeq 0.5-2.3$kHz.

\begin{figure}[htbp!]
\centering
\subfigure[Axisymmetric electrostatic potential]{
  \label{fig:phixy}
  \includegraphics[width=0.45\columnwidth]{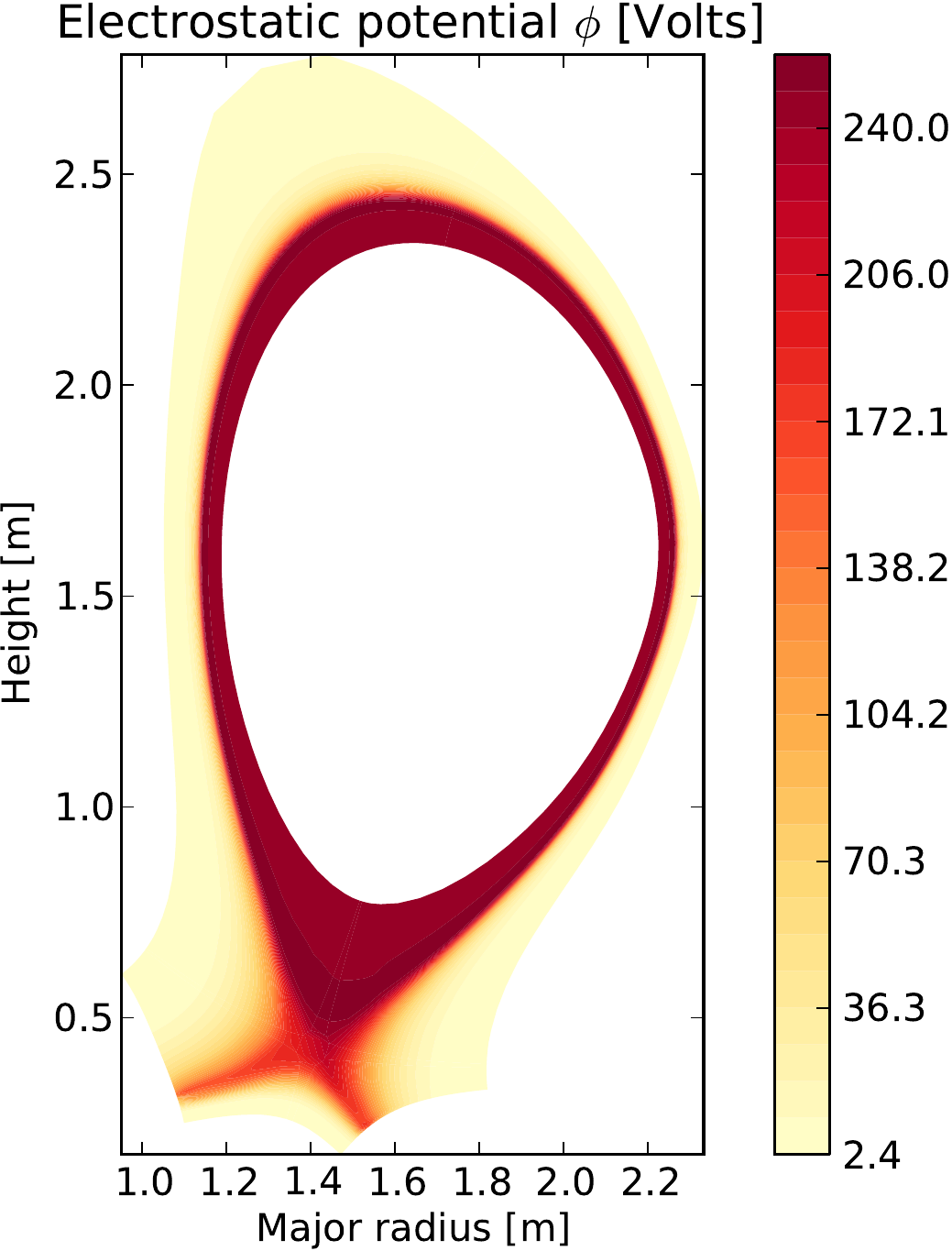}
}
\subfigure[Axisymmetric parallel current]{
  \label{fig:jparxy}
  \includegraphics[width=0.45\columnwidth]{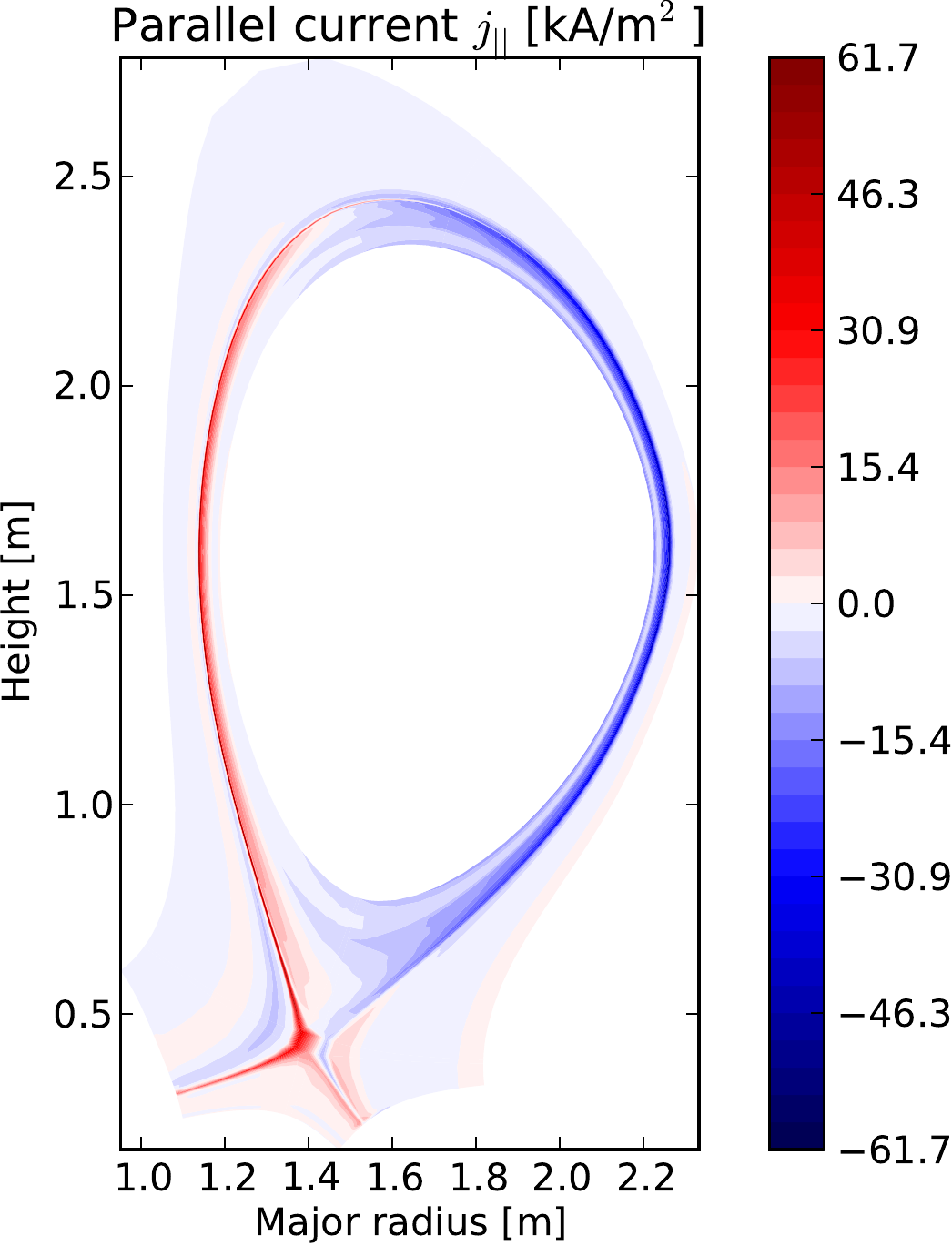}
}
\caption{Axisymmetric quasi-steady state electrostatic potential and parallel current after $t=0.83$ms, after initial transients shown in figure~\ref{fig:d3d_axisym_phi}.}
\label{fig:d3d_phij_xy}
\end{figure}

\begin{figure}
\centering
\includegraphics[width=0.6\columnwidth]{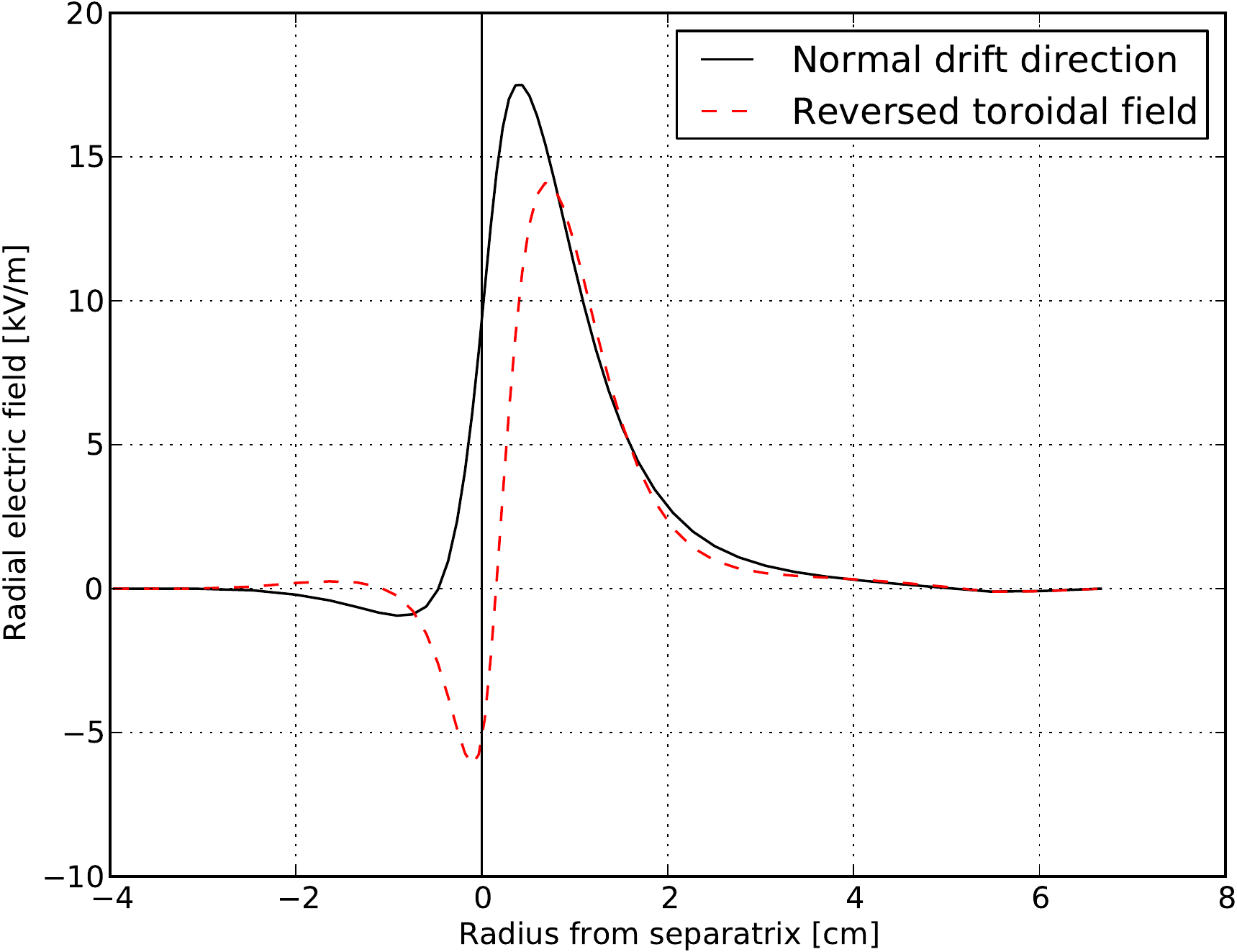}
\caption{Axisymmetric radial electric field at the outboard midplane at $t=0.83$ms (same time axis as figure~\ref{fig:d3d_phij_xy}) for normal (solid black) and reversed (dashed red) toroidal field direction. The normal configuration has the electron drift direction $\mathbf{V}_{mag}$ (equation~\ref{eq:drifts}) away from the active X-point (ion drift towards the X-point).}
\label{fig:d3d-midplane}
\end{figure}

The electrostatic potential after $t=0.83$ms on the time axis used in figure~\ref{fig:d3d_axisym_phi} (where currents are turned on at $t=0$) is shown in figure~\ref{fig:phixy}. The electrostatic potential is relatively constant in the closed field-line region, as shown in the plot of radial electric field in figure~\ref{fig:d3d-midplane}. This quasi-steady state is associated with a parallel current shown in figure~\ref{fig:jparxy}. The Pfirsch-Schl\"uter current pattern is seen in the closed field-line region, balancing the current due to the magnetic drift $\mathbf{V}_{mag}$ (diamagnetic current divergence). In the SOL and Private Flux (PF) region parallel currents are seen to flow into the sheath in steady state, requiring closing flows through the vessel walls. This is because in the boundary conditions (equation~\ref{eq:sheath_current}) we have assumed conducting walls. 
Also shown in figure~\ref{fig:d3d-midplane} is the radial electric field for a case with reversed toroidal field $B_\zeta$. This reverses the sign of the vertical magnetic drift $\mathbf{V}_{mag}$, leading to a modification of the radial electric field in the plasma edge. In the SOL the electric field is only slightly modified, since the sheath strongly constrains the electric field.


\subsection{Turbulence in X-point geometry}
\label{sec:xpt_turb}

Finally, once the axisymmetric electric field has reached quasi-steady state we turn off the imposed anomalous transport ($D_\perp$, $\chi_\perp$), extend the mesh in the toroidal domain to create a 3D simulation, and add small-scale random noise to the vorticity to seed instabilities. This is then run until it reaches a saturated turbulent state. 

\begin{figure}
\centering
\subfigure[Density fluctuations at the outboard midplane separatrix]{
  \label{fig:d3d-timeseries-ne}
  \includegraphics[width=0.45\columnwidth]{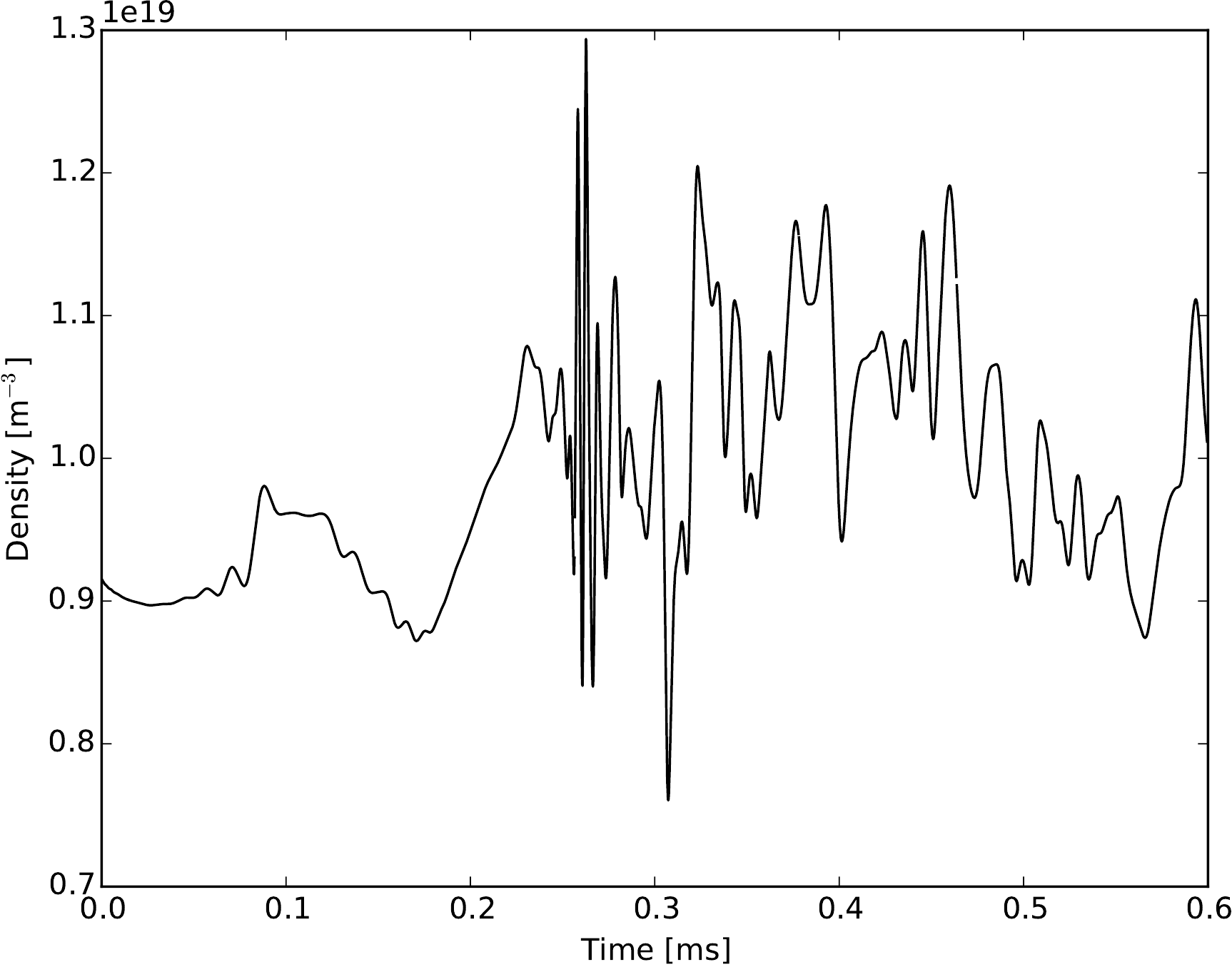}
}
\subfigure[Electrostatic potential at outboard midplane, 4cm inside separatrix]{
  \label{fig:d3d-timeseries-phi}
  \includegraphics[width=0.45\columnwidth]{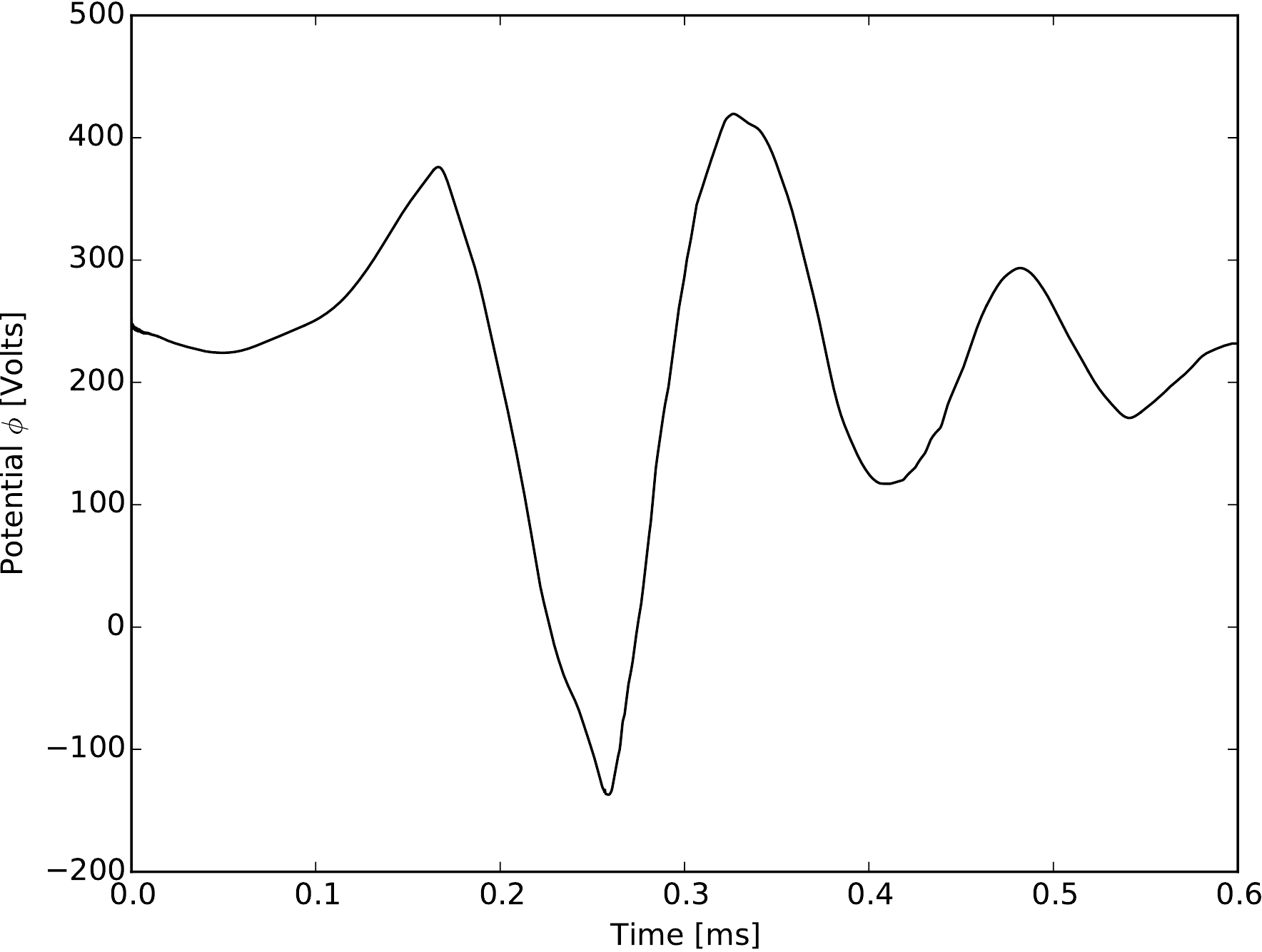}
}
\caption{Time series of density $n_e$ and potential $\phi$. The time axis begins at the start of the 3D simulation, initialised with the axisymmetric solution shown in figure~\ref{fig:d3d_phij_xy}}
\label{fig:d3d-timeseries-midplane}
\end{figure}

A timeseries of density fluctuations taken from the grid cell just outside the separatrix (which is located on a cell boundary) at the outboard midplane is shown in figure~\ref{fig:d3d-timeseries-midplane}. The density shows an initial drop, as the imposed anomalous diffusion has been turned off, and the plasma fluctuations start at small amplitude and so provide little transport initially.

\begin{figure}
\centering
\subfigure[Root Mean Square (RMS) density fluctuation amplitude]{
  \label{fig:d3d-rms-ne}
  \includegraphics[width=0.45\columnwidth]{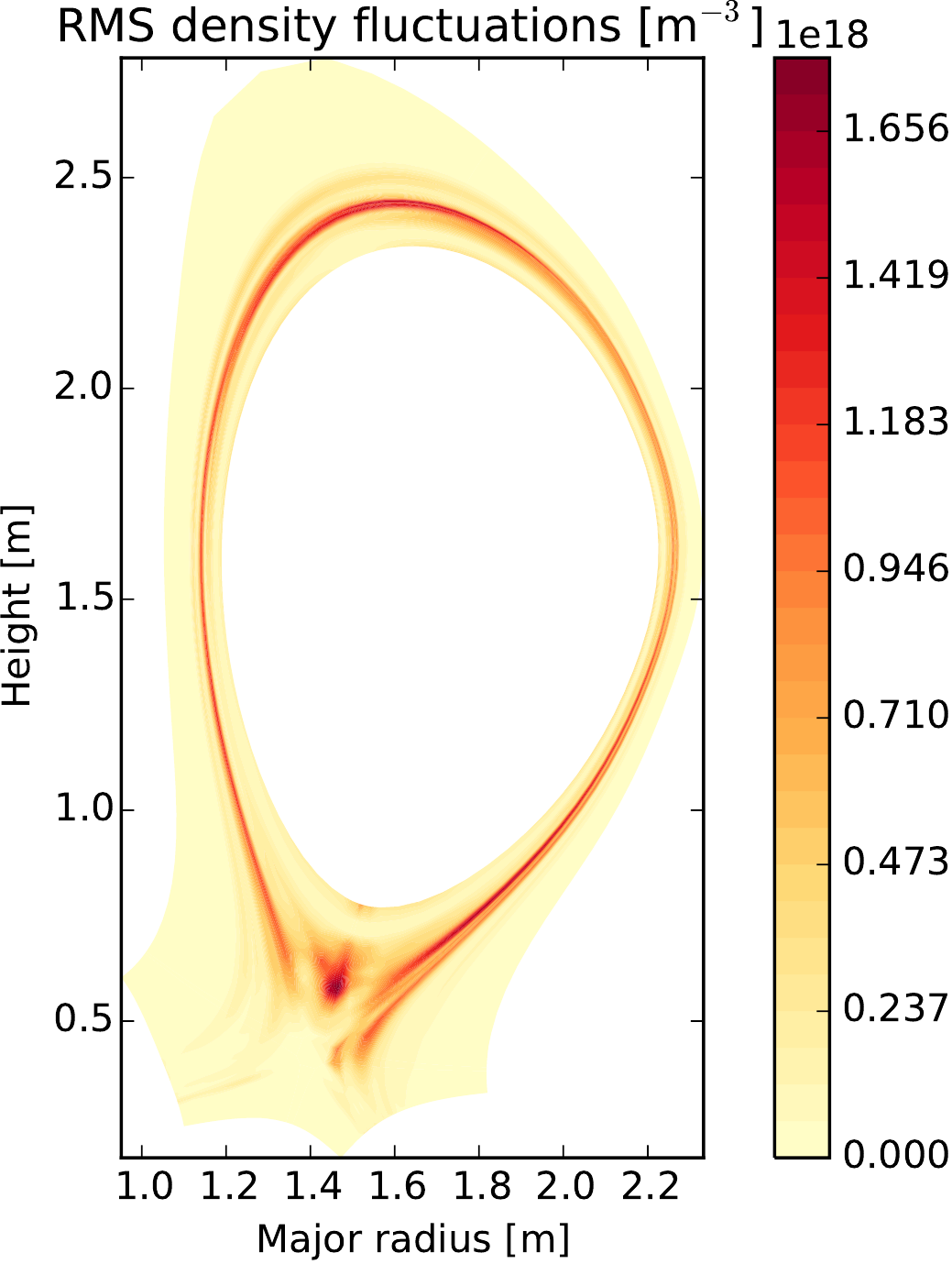}
}
\subfigure[Relative density fluctuations, as a percentage of the time and toroidal averaged density]{
  \label{fig:d3d-rms-ne-rel}
  \includegraphics[width=0.45\columnwidth]{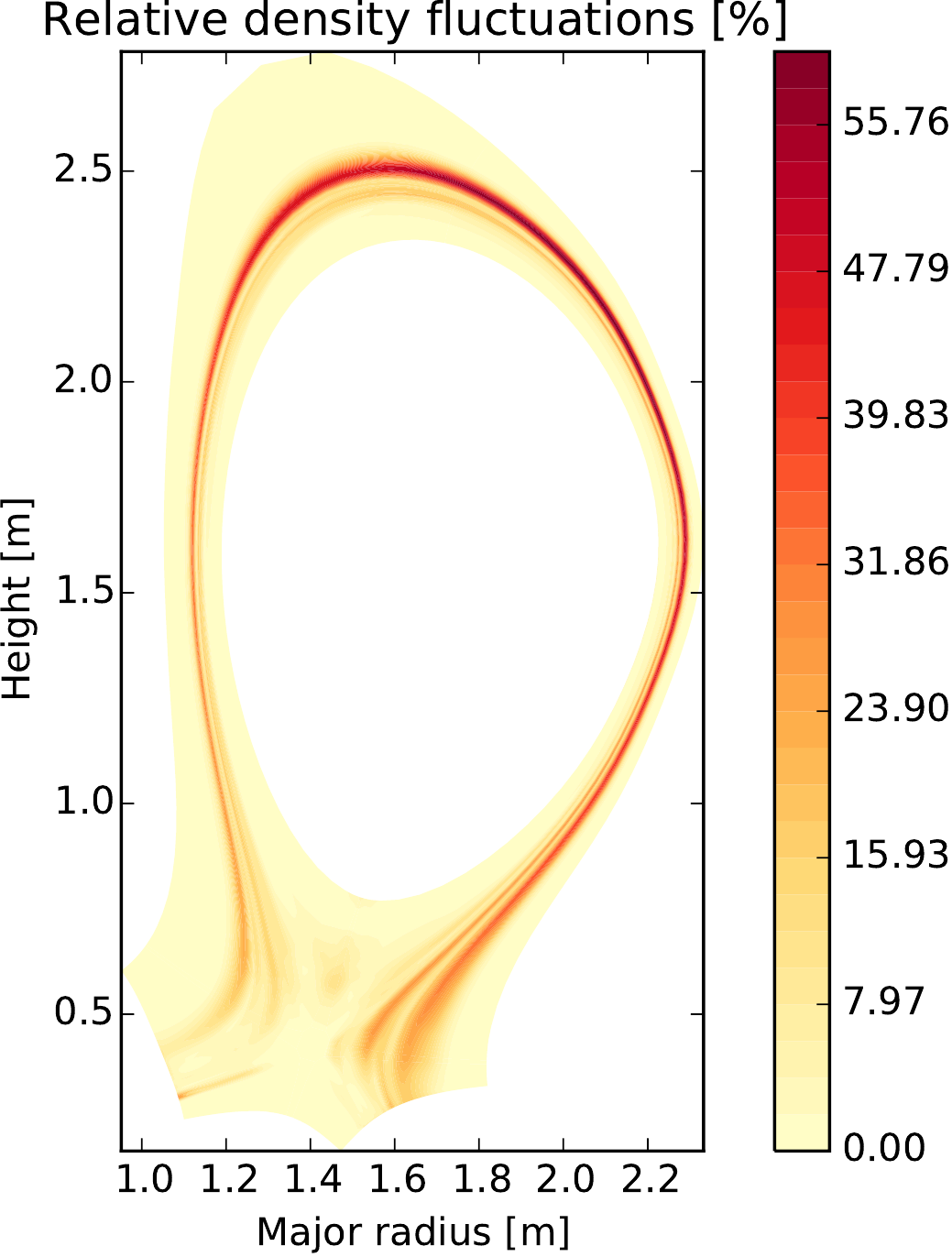}
}
\caption{Amplitude of electron density fluctuations in a simulation of DIII-D geometry, averaged over toroidal angle and the last $90\mu$s of simulation time (equation~\ref{eq:rms})}
\label{fig:d3d-rms-xy}
\end{figure}

Averaging over the toroidal angle and time, the Root Mean Square (RMS) density fluctuations
are calculated from the non-axisymmetric components of the density fluctuations (equation~\ref{eq:rms}).
\begin{equation}
n^{RMS} = \sqrt{\left< \left(n_e - \left<n_e\right>_\zeta\right)^2\right>_{\zeta, t}}
\label{eq:rms}
\end{equation}
where $\left<\cdot\right>_\zeta$ represents an average over toroidal angle $\zeta$, and $\left<\cdot\right>_{\zeta,t}$
an average over toroidal angle and time. This is shown in figure~\ref{fig:d3d-rms-xy}, showing a ballooning
character to the fluctuations, but with significant fluctuations at all poloidal angles close to the separatrix.
As can be more clearly seen in figure~\ref{fig:d3d-rms-ne-rel}, relatively large fluctuations are observed in the private flux region,
in particular in the bad curvature region of the inner leg, close to the divertor plate. This feature has been predicted in gradient-driven simulations~\cite{umansky-2005} and observed experimentally~\cite{harrison-2015}. The drive for these fluctuations is probably a combination of interchange drive due to bad curvature in this region, and a conducting wall mode due to the sheath~\cite{ryutov-2004}, though more detailed analysis is needed to confirm this.

Further analysis of longer timeseries will be the subject of future work. Here we have focussed on testing the ability of the Hermes model to simulate turbulent transport together with evolving profiles and electric fields. Since the anomalous transport coefficients chosen initially are unlikely to result in the same transport as the turbulence, the plasma profiles will continue to evolve. Long turbulent simulations are therefore still required to find a self-consistent solution, but the procedure described here greatly reduces the computational cost: The turbulence simulations (figure~\ref{fig:d3d-timeseries-midplane}) required approximately $7\times 10^5$ core-hours (on Archer) per millisecond of simulated time, whereas the fluid simulations (figure~\ref{fig:d3d_fluid_ne}) required approximately 16 core-hours per millisecond.

\section{Conclusions}
\label{sec:conclusions}

We have described the key features of Hermes, a new model based on BOUT++, which is being developed to study transport in the edge of magnetically confined plasmas, in particular tokamaks. Progress has been made towards self-consistent simulation of turbulence and profiles on transport timescales: a drift-reduced model has been chosen and analysed for its conservation properties; numerical methods have been developed which conserve integral properties of the analytical model and so allow long-time simulation of plasma oscillations; and a new method for solving the electrostatic potential in X-point geometry with a field-aligned coordinate system has been developed and implemented. Together, these improvements enable simulations to be carried out in poloidal limiter (ISTTOK) and diverted tokamak configurations, which self-consistently evolve the large-scale electric fields and currents alongside the turbulence. 
   Significant work remains to be done, some of which has been discussed in previous sections. Verification and validation of such a complex model will take significant effort, and is ongoing. As part of this work, we plan to carry out comparisons against ISTTOK in the near term as part of a EUROfusion project. The description of transport in the tokamak edge is strongly influenced by interaction with neutral gas, so in parallel with the work described here we have developed neutral gas models which will be described elsewhere. Hot ion effects, primarily ion diamagnetic drift and parallel viscosity, will modify the results presented here. Including these effects introduces complications, particularly in the vorticity equation, but is currently under developement. Finally, the description of the radial electric field (poloidal flow) in tokamak plasmas is a complex and subtle topic, and the present treatment will need further refinement. 

\section*{Acknowledgements}

This work has been carried out within the framework of the EUROfusion Consortium and has received funding from the Euratom research and training programme 2014-2018 under grant agreement No 633053. The views and opinions expressed herein do not necessarily reflect those of the European Commission. The authors gratefully acknowledge the support of the UK Engineering and Physical Sciences Research Council (EPSRC) under grant EP/K006940/1, and Archer computing resources under Plasma HEC consortium grant EP/L000237/1.

\section*{References}

\bibliography{hermes}

\begin{thebibliography}{10}

\bibitem{kukushkin2011}
A~S Kukushkin, H~D Pacher, V~Kotov, G~W Pacher, and D~Reiter.
\newblock {\em Fusion Engineering and Design}, 86:2865--2873, 2011.

\bibitem{efdaroadmap}
F~Romanelli et~al.
\newblock {Fusion Electricity - A roadmap to the realisation of fusion energy}.
\newblock Technical report, {EFDA}, 2012.

\bibitem{wennninger2015}
R~Wenninger, F~Arbeiter, J~Aubert, L~Aho-Mantila, R~Albanese, R~Ambrosino,
  C~Angioni, J.-F Artaud, M~Bernert, E~Fable, A~Fasoli, G~Federici, J~Garcia,
  G~Giruzzi, F~Jenko, P~Maget, M~Mattei, F~Maviglia, E~Poli, G~Ramogida,
  C~Reux, M~Schneider, B~Sieglin, F~Villone, M~Wischmeier, and H~Zohm.
\newblock 55:063003, 2015.

\bibitem{naulin-2007}
V~Naulin.
\newblock {\em J. Nucl. Materials}, 363-365:24--31, 2007.

\bibitem{ghendrih-2012}
Ph~Ghendrih, C~Norscini, F~Hasenbeck, G~Dif-Pradalier, J~Abiteboul,
  T~Cartier-Michaud, X~Garbet, V~Grandgirard, Y~Marandet, and Y~Sarazin.
\newblock {\em J.Phys.: Conf. Ser.}, 401:012007, 2012.

\bibitem{dippolito-2011}
D~A D'Ippolito, J~R Myra, and S~J Zweben.
\newblock {\em Physics of Plasmas}, 18:060501, 2011.

\bibitem{ricci2012}
P~Ricci, F~D Halpern, S~Jolliet, J~Loizu, A~Mosetto, A~Fasoli, I~Furno, and
  C~Theiler.
\newblock {\em Plasma Phys. Control. Fusion}, 54:124047, 2012.

\bibitem{halpern2016}
F~D Halpern, P~Ricci, S~Jolliet, J~Loizu, J~Morales, A~Mosetto, F~Musil,
  F~Riva, T~M Tran, and C~Wersal.
\newblock {\em J. Comput. Phys.}, 315:388--408, 2016.

\bibitem{tamain2010}
P~Tamain, Ph~Ghendrih, E~Tsitrone, V~Grandgirard, X~Garbet, Y~Sarazin, E~Serre,
  G~Ciraolo, and G~Chiavassa.
\newblock {\em J. Comput. Phys.}, 229(2):361--378, 2010.

\bibitem{tamain2016}
P~Tamain, H~Bufferand, G~Ciraolo, C~Colin, D~Galassi, Ph. Ghendrih,
  F~Schwander, and E~Serre.
\newblock {\em J. Comput. Phys.}, In Press, 2016.

\bibitem{catto-2008}
P~J Catto, A~N Simakov, F~I Parra, and G~Kagan.
\newblock {\em Plasma Phys. Control. Fusion}, 50:115006, 2008.

\bibitem{mikhailovskii1984}
A~B Mikhailovskii and V~S Tsypin.
\newblock {\em Beitr. Plasmaphys.}, 24:335–354, 1984.

\bibitem{pfirsch1996}
D~Pfirsch and D~Correa-Restrepo.
\newblock {\em Plasma Phys. Control. Fusion}, 38:71--101, 1996.

\bibitem{mikhailovskii1997}
A~B Mikhailovskii et~al.
\newblock {\em Plasma Phys. Rep.}, 23:844, 1997.

\bibitem{reiser2012}
D~Reiser.
\newblock {\em Physics of Plasmas}, 19:072317, 2012.

\bibitem{simakov-2003}
A~N Simakov and P~J Catto.
\newblock {\em Physics of Plasmas}, 10(12):pp. 4744--4757, December 2003.

\bibitem{scott-2003}
B~Scott.
\newblock {\em Physics of Plasmas}, 10:963, 2003.

\bibitem{Dudson2009}
B~D Dudson et~al.
\newblock {\em Comp. Phys. Comm.}, 180:1467--1480, 2009.

\bibitem{dudson2014}
B~D Dudson et~al.
\newblock 81(01):365810104, 2015.
\newblock doi:10.1017/S0022377814000816.

\bibitem{petsc-web-page}
Satish Balay, Shrirang Abhyankar, Mark~F. Adams, Jed Brown, Peter Brune, Kris
  Buschelman, Lisandro Dalcin, Victor Eijkhout, William~D. Gropp, Dinesh
  Kaushik, Matthew~G. Knepley, Lois~Curfman McInnes, Karl Rupp, Barry~F. Smith,
  Stefano Zampini, Hong Zhang, and Hong Zhang.
\newblock {PETS}c {W}eb page.
\newblock http://www.mcs.anl.gov/petsc, 2016.

\bibitem{petsc-user-ref}
Satish Balay, Shrirang Abhyankar, Mark~F. Adams, Jed Brown, Peter Brune, Kris
  Buschelman, Lisandro Dalcin, Victor Eijkhout, William~D. Gropp, Dinesh
  Kaushik, Matthew~G. Knepley, Lois~Curfman McInnes, Karl Rupp, Barry~F. Smith,
  Stefano Zampini, Hong Zhang, and Hong Zhang.
\newblock {PETS}c users manual.
\newblock Technical Report ANL-95/11 - Revision 3.7, Argonne National
  Laboratory, 2016.

\bibitem{petsc-efficient}
Satish Balay, William~D. Gropp, Lois~Curfman McInnes, and Barry~F. Smith.
\newblock Efficient management of parallelism in object oriented numerical
  software libraries.
\newblock In E.~Arge, A.~M. Bruaset, and H.~P. Langtangen, editors, {\em Modern
  Software Tools in Scientific Computing}, pages 163--202. Birkh{\"{a}}user
  Press, 1997.

\bibitem{winsor-1968}
N~Winsor, J~L Johnson, and J~M Dawson.
\newblock {\em Phys. Fluids}, 11:2448, 1968.

\bibitem{silva-2004}
C~Silva, I~Nedzelskiy, H~Figueireda, R~M~O Galvao, J~A~C Cabral, and C~A~F
  Varandas.
\newblock {\em Nucl. Fusion}, 44:799--810, 2004.

\bibitem{schneider-2006}
R~Schneider, X~Bonnin, K~Borrass, D~P Coster, H~Kastelewicz, D~Rieter, V~A
  Rozhansky, and B~J Braams.
\newblock {\em Contrib. Plasma Phys.}, 46(1-2):3--191, 2006.

\bibitem{angus2014}
J~R Angus and M~V Umansky.
\newblock {\em Physics of Plasmas}, 21:012514, 2014.

\bibitem{omotani2016}
J~T Omotani, F~Militello, L~Easy, and N~Walkden.
\newblock {\em Plasma Phys. Control. Fusion}, 58:014030, 2016.

\bibitem{scott-2005}
B~Scott.
\newblock {\em Physics of Plasmas}, 12:102307, 2005.

\bibitem{chodura-1982}
R~Chodura.
\newblock {\em Phys. Fluids}, 25:1628, 1982.

\bibitem{stangeby-2000}
P~C Stangeby.
\newblock {\em {The Plasma Boundary of Magnetic Fusion Devices}}.
\newblock IoP, 2000.

\bibitem{loizu-2012}
J~Loizu, P~Ricci, F~D Halpern, and S~Jolliet.
\newblock {\em Physics of Plasmas}, 19:122307, 2012.

\bibitem{siddiqui-2015}
M~U Siddiqui, D~S Thompson, C~D Jackson, J~F Kim, N~Hershkowitz, and E~E Scime.
\newblock {\em Physics of Plasmas}, 23:057101, 2016.

\bibitem{togo-2016}
S~Togo, T~Takizuka, M~Nakamura, K~Hoshino, K~Ibano, T~Long~Lang, and Y~Ogawa.
\newblock {\em J. Comput. Phys.}, 310:109--126, 2016.

\bibitem{ghendrih-2011}
Ph~Ghendrih, K~Bodi, H~Bufferand, G~Chiavassa, G~Ciraolo, N~Fedorczak,
  L~Isoardi, A~Paredes, Y~Sarazin, and E~Serre.
\newblock {\em Plasma Phys. Control. Fusion}, 53(5):054019, 2011.

\bibitem{bufferand-2014}
H~Bufferand, G~Ciraolo, G~Dif-Pradalier, P~Ghendrih, Ph~Tamain, Y~Marandet, and
  E~Serre.
\newblock {\em Plasma Phys. Control. Fusion}, 56:122001, 2014.

\bibitem{peterson2013}
J~L Peterson and G~W Hammett.
\newblock {\em {SIAM} J. Sci. Computing}, 35(B576), 2013.

\bibitem{hindmarsh2005}
A~C Hindmarsh et~al.
\newblock {\em {ACM} Transactions on Mathematical Software}, 31(3):363--396,
  2005.

\bibitem{zhou-2015}
D~Zhou.
\newblock {\em Physics of Plasmas}, 22:092504, 2015.

\bibitem{rognlien-1999}
T~D Rognlien, D~D Ryutov, N~Mattor, and G~D Porter.
\newblock {\em Physics of Plasmas}, 6:1851, 1999.

\bibitem{chankin-2000}
A~V Chankin, J~P Coad, G~Corrigan, S~J Davies, S~K Erents, H~Y Guo, G~F
  Matthews, G~J Radford, J~Spence, P~C Stangeby, and A~Taroni.
\newblock {\em Contrib. Plasma Phys.}, 40(3-4):288--294, 2000.

\bibitem{chankin-2015}
A~V Chankin, G~Corrigan, M~Groth, P~C Stangeby, and {JET contributors}.
\newblock {\em Plasma Phys. Control. Fusion}, 57:095002, 2015.

\bibitem{arakawa-1966}
A~Arakawa.
\newblock {\em J. Comput. Phys.}, 1(1):119--143, 1966.

\bibitem{murthy-2002}
J~Y Murthy.
\newblock Numerical methods in heat, mass and momentum transfer.
\newblock Lecture notes ME608, 2002, Purdue University.

\bibitem{xu-2008}
X~Q Xu, M~V Umansky, B~Dudson, and P~B Snyder.
\newblock {\em Comm. in Comput. Phys.}, 4(5):pp. 949--979, November 2008.

\bibitem{dimits-1993}
A~M Dimits.
\newblock {\em Phys. Rev. E}, 48(5):4070--4079, Nov 1993.

\bibitem{scott01}
B~Scott.
\newblock {\em Physics of Plasmas}, 8(2):447, 2001.

\bibitem{rognlien-2002}
T~D Rognlien, X~Q Xu, and A~C Hindmarsh.
\newblock {\em J. Comput. Phys.}, 175:249--268, 2002.

\bibitem{umansky-2005}
M~V Umansky, T~D Rognlien, and X~Q Xu.
\newblock {\em J. Nucl. Materials}, 337-339:266--270, 2005.

\bibitem{harrison-2015}
J~R Harrison, G~M Fishpool, and B~D Dudson.
\newblock {\em J. Nucl. Materials}, 463:757--760, 2015.

\bibitem{ryutov-2004}
D~D Ryutov and R~H Cohen.
\newblock {\em Contrib. Plasma Phys.}, 44(1-3):168–175, 2004.

\end{thebibliography}
\bibliographystyle{unsrt}

\end{document}